%% file: main_rev.tex
\documentclass{article}

\usepackage[english]{babel}

\usepackage[letterpaper,top=2cm,bottom=2cm,left=3cm,right=3cm,marginparwidth=1.75cm]{geometry}

\usepackage{amsmath}
\usepackage{graphicx}
\usepackage{xcolor}
\usepackage[colorlinks=true, allcolors=blue]{hyperref}
\usepackage{subcaption}
\usepackage{indentfirst}
\usepackage{booktabs}
\usepackage{nicefrac}
\usepackage{mdframed}
\usepackage{authblk}

\newenvironment{revision}
  {\color{red}\hypersetup{allcolors=red}}
  {\hypersetup{allcolors=blue}}

\title{Green building blocks reveal the complex anatomy of climate change mitigation technologies}
\author[1,2]{Yang Li}
\author[3,4]{Frank Neffke}

\affil[1]{School of Resources and Environment, Nanchang University. No. 999 Xuefu Avenue, Nanchang 330031, Jiangxi Province, China}
\affil[2]{The Growth Lab, Harvard University, 79 JFK Street, Cambridge
 02138, MA, USA}
\affil[3]{Complexity Science Hub. Metternichgasse 8, 1030 Vienna, Austria}
\affil[4]{Transforming Economies Lab, IT:U Interdisciplinary Transformation University Austria, Altenberger Straße 66c, 4040 Linz, Austria}
\begin{document}
\maketitle

\begin{abstract}

Achieving net-zero emissions requires rapid innovation, yet the necessary technological knowhow is scattered across industries and countries. Comparing functionally similar green and nongreen patents, we identify ``Green Building Blocks'' (GBBs): modular components that can be added to reduce  existing technologies' carbon footprints. These GBBs depict the anatomy of the green transition as a network that connects problems --- nongreen technologies --- to GBBs that mitigate their climate-change impact. Node degrees in this network are highly unequal, showing that the scope for climate-change mitigating innovation varies substantially across domains. \textcolor{red}{The network helps predict which green technologies firms develop and maps which organizations hold complementary capabilities.} \begin{revision}The most complementary potential partners reside abroad for 84\% of US firms, 87\% of German firms, and 92\% of Chinese firms, highlighting the importance of being able to collaborate across national borders.\end{revision}
\end{abstract}

\section*{Introduction}

Achieving carbon neutrality by 2050 requires a radical transformation of how the global economy produces the products and services we consume. While there is broad agreement on the pressing need to reduce anthropogenic carbon emissions and on the ultimate goal of carbon neutrality \cite{intergovernmental_panel_on_climate_change_ipcc_climate_2023}, which technological trajectories hold most promise for reaching this goal  remains debated.
One challenge is that there is no coherent set of  ``green'' technologies. Different economic sectors require different solutions, from transforming emission-intensive manufacturing processes (e.g., steel\cite{xu_plant-by-plant_2023,kazmi_state---art_2023} or cement\cite{habert_environmental_2020,nehdi_is_2024}) to upgrading the efficiency of buildings\cite{camarasa_global_2022, leibowicz_optimal_2018} and overhauling transportation systems\cite{zhang_cross-cutting_2022, melton_moving_2016}.  This inherent diversity suggests that the green transition is no monolithic shift in technological paradigm \cite{freeman_structural_1988} but a complex tapestry of custom tailored innovations.

However, we know little about the synergies across domains in terms of transferability of technological solutions. To remedy this, we map  how technological solutions can be reused across different domains of application. To do so, we conceptualize the space of technologies as consisting of \emph{nongreen sources} or problems to be solved,  \emph{green building blocks} (GBBs) --- which can be added to these sources to reduce or eliminate their  carbon emissions --- and the specific \emph{green solutions} that emerge as a result. This recasts the green transition as a network that connects nongreen source technologies to GBBs. \begin{revision}This network identifies how component capabilities can be reused across technological domains and which source fields have a narrower set of options available to mitigate climate change impact, because their green inventions draw on comparatively few GBBs.\end{revision} Furthermore, we use this network to describe firms'  climate-change mitigation capabilities in terms of their GBBs, to predict the new green solutions these firms develop, and \textcolor{red}{to identify which candidate partners could offer complementary capabilities}.

Many scholars subscribe to the idea that innovation is an important tool in our arsenal to combat climate change. Existing research has produced a wealth of micro-level carbon-reduction processes and technologies, often described in detailed engineering case studies and life cycle assessments\cite{zoback_meeting_2023, arshad_life_2022}. While invaluable, these studies predominantly focus on solutions within specific sectors, without exploring how they connect across domains of application. On the other end of the spectrum, integrated assessment models (IAMs) sketch macro-level trajectories and system-wide scenarios that serve as road maps to a net zero economy \cite{davis_net-zero_2018,speizer_integrated_2024}. \begin{revision}Our approach complements these with an empirical map of fine-grained, technological components ---GBBs--- and their distribution across firms and countries.\end{revision}

\begin{revision}We develop this complementary perspective\end{revision} by drawing from theories of combinatorial innovation\cite{schumpeter_theory_1934, weitzman_recombinant_1998, arthur_nature_2009, youn_invention_2015}. We hypothesize that the green transition requires combining existing technologies with GBBs such that new, climate-change mitigating innovations emerge. That is, rather than focusing on the green technologies themselves, we view these technologies as consisting of modular components, some of which can be reused to reduce carbon emissions across technological domains. We illustrate this logic in Fig.~\ref{fig:fig1}, which uses a 2-dimensional embedding of Cooperative Patent Classification (CPC) classes --- curated labels that describe the  technologies related to patented inventions --- to depict nongreen sources (on the left), GBBs (in the center) and climate-change-mitigation solutions (CCMTs, on the right). \begin{revision}Throughout, CCMTs refer to patented technologies identified by CPC Y02 (``Technologies or applications for mitigation or adaptation against climate change'') and Y04 (``Information or communication technologies having an impact on other technological areas'', which includes smart grids) labels. Our analysis focuses on modifications of existing technical functions. Systemic transitions and nonpatented organizational or business-model changes are beyond our scope.\end{revision}

We identify GBBs from patent data. We first pair nongreen patents to CCMT patents that fulfill similar functions based on their textual similarity. CCMT patents in such pairings typically list more CPC codes than their nongreen counterparts. We focus on these ``extra'' CPC codes and extract clusters of CPC codes that often co-occur. This yields a set of 82 coherent sets of technology codes used in such ``additions,'' which  we interpret as  GBBs. GBBs span a variety of plausible technological approaches to climate-change mitigation. For instance, we identify efficiency enhancing GBBs, such as \emph{Advanced Heating Solutions}, GBBs that leverage new materials, such as \emph{Membrane Technologies}, and GBBs that support electrification, such as \emph{Electrical Coupling Devices}. \begin{revision}Our building blocks are ``green'', not because of some intrinsic characteristics, but due to the role they play in technological combinations: GBBs are components that we empirically observe in climate-mitigating implementations of a given technological function but not in its nongreen implementations.\end{revision}

Our analysis yields three key findings. First, the network that connects nongreen sources to GBBs has a nontrivial topology: whereas some GBBs connect to relatively few source fields, representing  specialized solutions for specific problems, other GBBs are general-purpose technologies that can be applied across a wide range of domains. Second, at the micro-level of corporate innovation behavior, GBBs help predict which firms will develop which types of green inventions, offering a new way to assess firms' green technological capabilities. \begin{revision}Third, GBBs identify which organizations hold capabilities that could help a firm mitigate its climate change impact, but that are missing from its own capability portfolio. The most complementary candidates often reside abroad, revealing the relevance of international cooperation in finding technological solutions for emission-intensive activities.\end{revision}

Our research builds upon and extends several existing research streams. It complements efforts to classify economic activities as green or brown\cite{mealy_economic_2022,andres_stranded_2023}, starting from a combinatorial view of innovation. It also relates to studies that extract technological pathways from patent citations\cite{linares_patent-based_2019, zhou_patent_2014}, drawing inspiration from research on techno-economic paradigms \cite{dosi_technological_1982, freeman_structural_1988}. However, our emphasis on the modularity and cross-sectoral applicability of GBBs provides an alternative perspective, shifting the focus from retracing technological trajectories\cite{dosi_technological_1982} by analyzing historical citation trees \cite{verspagen_mapping_2007,nomaler_patent_nodate} or co-occurrence patterns \cite{alstott_mapping_2017,stojkoski_multidimensional_2023} to the identification of green building blocks that can be reused in novel ways \begin{revision}across technological domains. A related literature, reviewed in \cite{caldarola2024economic}, applies relatedness and economic-complexity measures to green technologies, mostly at the level of regions and countries. We differ by identifying the components that are added in green inventions and by studying how they shape the innovation and collaboration of firms. These firm-level analyses connect the GBB map to organizational questions around diversification\cite{penrose1959theory,teece1982towards,farjoun1994beyond} and collaboration\cite{mowery1998technological,nooteboom2007optimal}, including work showing that firms rely on external knowledge and cooperation for environmental innovation\cite{ghisetti2015open}. In this context, GBBs offer a new way to assess firms' green capabilities and how they are reflected in strategic decisions.\end{revision}

\section*{Results}

\subsection*{\textcolor{red}{Identifying green building blocks}}

\begin{figure}[htp]
\centering
\begingroup
\phantomsubcaption\label{fig:fig1a}%
\phantomsubcaption\label{fig:fig1b}%
\phantomsubcaption\label{fig:fig1c}%
\endgroup
\includegraphics[width=\linewidth]{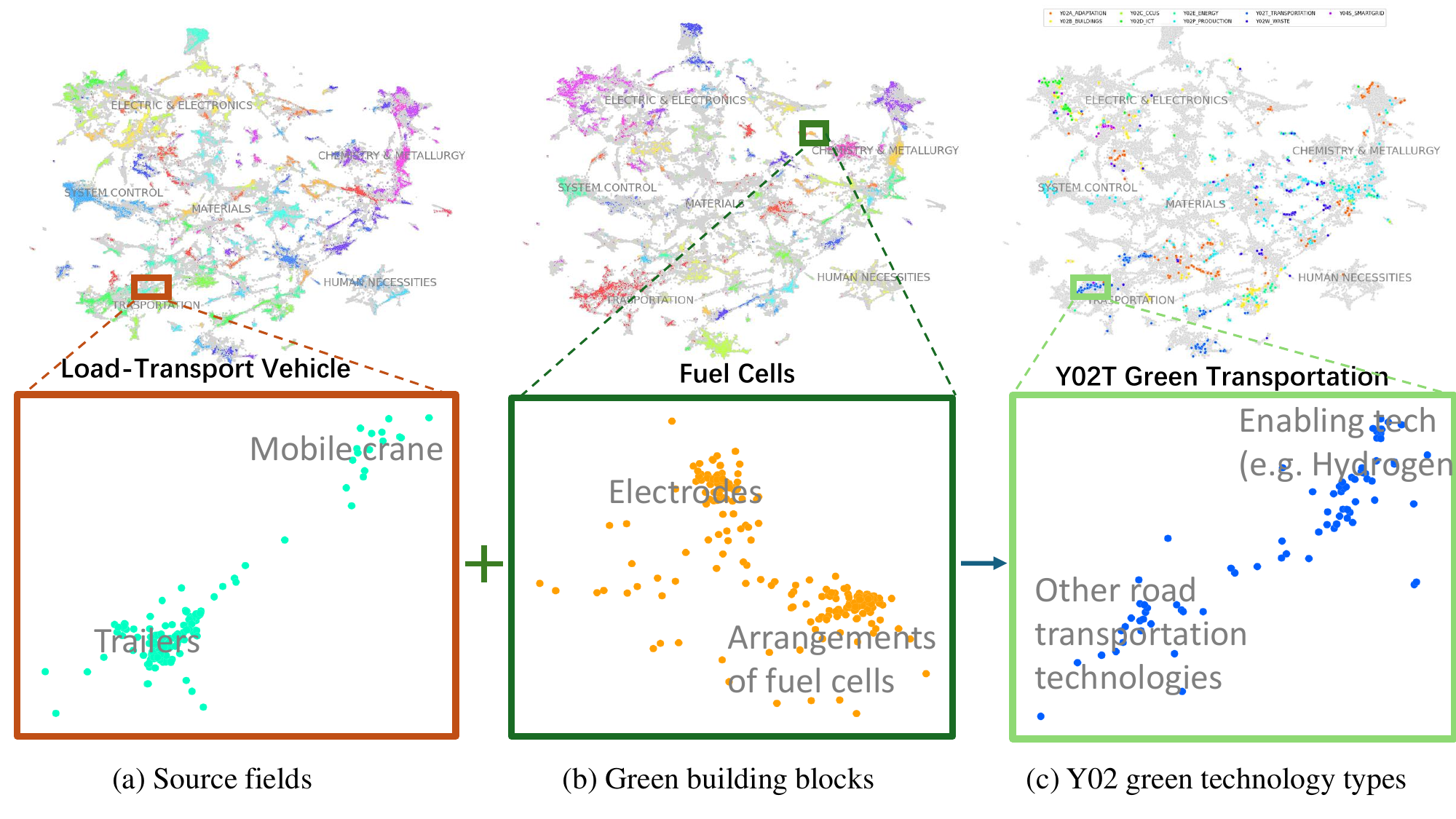}
\caption{The anatomy of CCMT innovation: source fields, green building blocks (GBBs), and CCMT technologies. Each point in the figure represents a CPC technology class plotted in a 2-dimensional layout generated by UMAP. Colors highlight CPC codes belonging to (\subref{fig:fig1a}) different sources fields, (\subref{fig:fig1b}) GBBs, or (\subref{fig:fig1c}) CCMT technologies. All other nodes are kept grey. The second row shows an example where \emph{load-transport vehicles} are combined with \emph{fuel cells} to produce green transportation related innovation. \textcolor{red}{This illustrates the paper's core decomposition: a source field representing a certain function, a recurrent additional component, and the resulting climate-mitigating implementation.} An interactive version of this visualization is available at \url{https://complexly.github.io/gbb/}.}
\label{fig:fig1}
\end{figure}

Green patents on average list more CPC codes than nongreen patents (see SI, Fig.~\ref{fig:morecodes}), especially when comparing green and nongreen technologies within the same broad technological fields (see SI, Table~\ref{tab:paired_diff}). \begin{revision}This suggests that green technologies are created by adding specific types of components to existing, nongreen technologies, thereby modifying these existing technologies.\end{revision} To be sure, the greater number of CPC codes may also reflect differences in technological breadth or classification practices. Our analysis therefore does not rely on the number of codes, but on which codes are added relative to functionally similar nongreen patents. Specifically, we propose that green technologies often consist of two types of components. The first is a traditional, nongreen, technology that aims to fulfill a particular function. For instance, the traditional \emph{heavy-load vehicles}, such as trucks, of Fig.~\ref{fig:fig1a} fulfill the function of transporting goods. The second type of components are GBBs, technological modules that help reduce carbon emissions, such as the \emph{fuel cell} technologies highlighted in Fig.~\ref{fig:fig1b}. When the two components are combined, they result in heavy-load vehicles that use fuel cells for their propulsion, producing  transportation that avoids carbon emissions (Fig.~\ref{fig:fig1c}).

\begin{figure}[htp]
\centering
\begingroup
\phantomsubcaption\label{fig:fig2a}%
\phantomsubcaption\label{fig:fig2b}%
\phantomsubcaption\label{fig:fig2c}%
\endgroup
\includegraphics[width=0.95\linewidth]{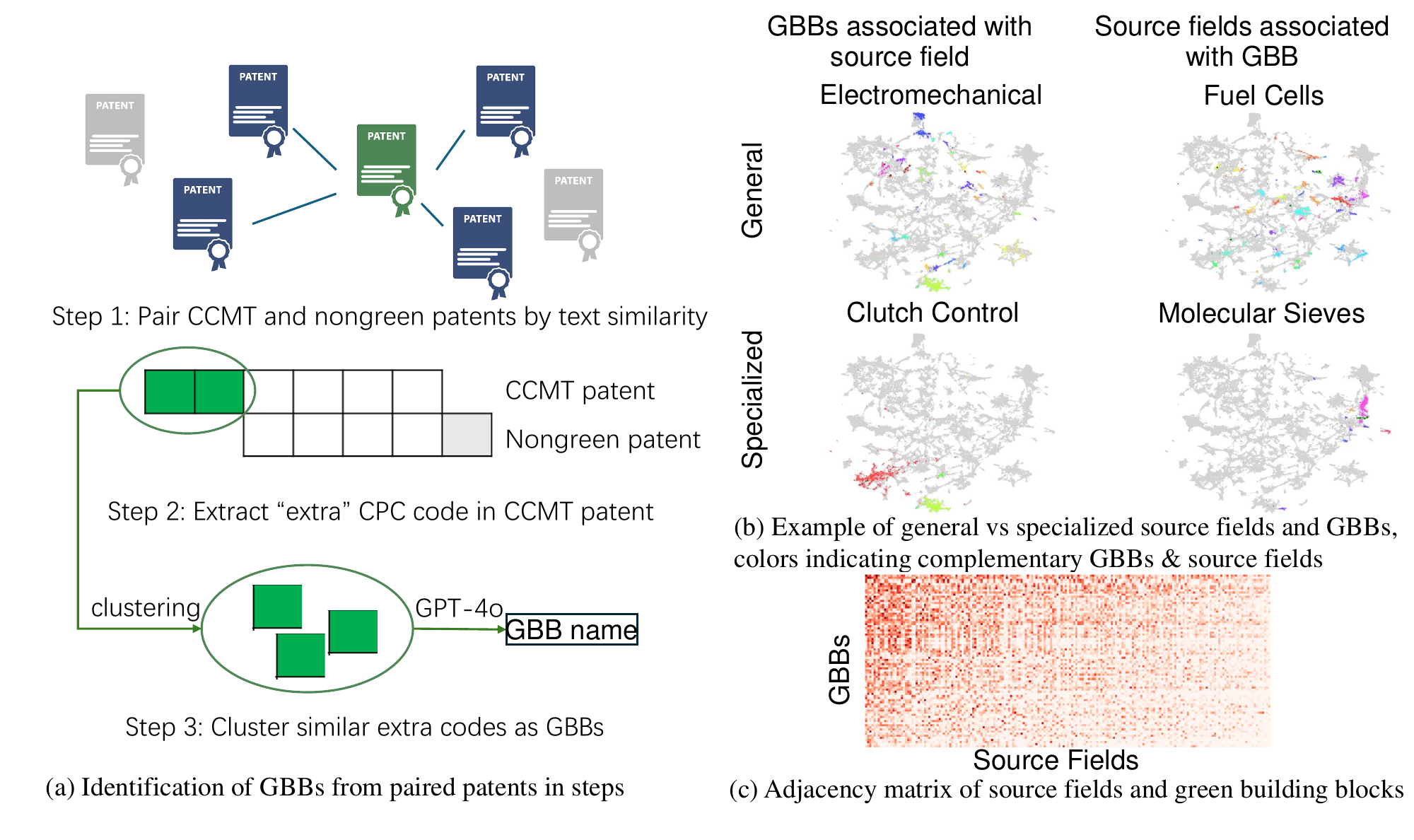}
\caption{Generality and specificity of source fields and green building blocks. \textbf{(\subref{fig:fig2a})}: Three-step identification of GBBs: (1) match CCMT to nongreen patents, (2) identify ``extra'' technology codes on the CCMT patent, (3) cluster extra codes into GBBs, and name GBBs with large language model. \textbf{(\subref{fig:fig2b})}: Examples of link between GBBs and source fields in 2-by-2 matrix depicting  general/specialized GBBs/source fields. \textbf{(\subref{fig:fig2c})}: Adjacency matrix of bipartite network connecting GBBs to source fields. \textcolor{red}{The triangular structure of this matrix shows that the number of observed options available to source fields, as well as the number of source fields GBBs serve, vary widely.}}
\label{fig:fig2}
\end{figure}

How can one identify GBBs? Our approach is laid out in Fig.~\ref{fig:fig2a}. We start by identifying ``green'' patents, selecting all patents that carry a Y02/04 technology code, which the patent office uses to identify climate-change mitigation technologies\cite{angelucci_supporting_2018}. Note that the Y02 class also covers technologies that aim at adaptation to climate change. As these make up only a few of our CCMT classes, we subsume them under the mitigation label. Next, we pair each CCMT patent to up to ten nongreen patents that aim to perform similar functions. To do so, we use sentence embeddings to find the most similar nongreen patents in terms of the function described in the patent text  (see Methods section). Of the 288,579 CCMT patent families, 239,862 have at least one valid nongreen match using FSE. We validate this approach, by comparing our ranking of functional similarity in patent pairs to human ratings. \textcolor{red}{Mean agreement with the human raters is 0.638, against an inter-rater agreement of 0.785 (see Methods and SI section \ref{sec:SI_vali_human}, which also replicates and extends this analysis using language-model based judges).}

Within each green-nongreen patent pair, we extract all codes on the CCMT patent that are not found on its matched nongreen counterpart, ignoring the Y02 and Y04 CCMT classes themselves. We regard these extra CPC codes as instantiations of green building blocks. A clustering algorithm (HDBSCAN) then groups the ``extra'' codes into green building blocks, and the common codes (shared by both CCMT and nongreen patents) into coherent source fields.

Due to noise in the attribution of  technology codes to patents, CCMT patents often list technology codes that are closely related, yet not exactly the same as those on their nongreen counterparts. Therefore, we denoise the signals derived from a patent's technology codes, by training a CPC2vec embedding to characterize the similarity of different CPC codes. Next, we define a \textcolor{red}{distance score} based on the embedding similarity. We use this score to \textcolor{red}{keep only those added codes that are sufficiently distinct from each of the codes on the matched nongreen patent. Finally, we aggregate the selected additions across all patent pairs}  (see Methods). In an alternative noise-reduction strategy, we  embed the entire patent in a lower dimensional technology space and extract GBBs by identifying the component of the low-dimensional vector describing the technology of a CCMT patent that is orthogonal to the vector describing its nongreen counterpart. In section~\ref{sec:SI_alternative_methods} of the SI, we show that this approach, as well as other variations, yield highly similar GBBs. \begin{revision}Moreover, the broad source--GBB structure persists under alternative text-embedding models, CPC resolutions, and clustering settings (see SI, sections \ref{sec:SI_gbb_stability} and \ref{sec:SI_cpc_granularity}).\end{revision}

The end result is a set of 82 green building blocks, as well as 193 source fields of nongreen technologies with which these GBBs can be combined. Note that GBBs and source fields are essentially collections of detailed CPC codes. We label both types of collections ---GBBs and source fields--- using a large language model, GPT-4o (see Methods). From hereon, we will use these labels to refer to specific source fields or GBBs. 

\subsection*{\textcolor{red}{Generality and specificity of source fields and green building blocks}}

Fig.~\ref{fig:fig2b} provides concrete examples that illustrate the relation between source fields and GBBs. The left column highlights GBBs for two different source fields, one in each row. The source field of \emph{electromechanical technologies} in the top connects to many different GBBs, opening up a variety of decarbonization strategies. In contrast, \emph{clutch-control technologies} in the bottom have only few available routes to decarbonization. The second column, instead, focuses on GBBs and plots the source fields   to which they can be applied, showing that --- as one would expect --- \emph{fuel cells} have a much wider range of application than \emph{molecular sieves}. These examples reflect broader patterns. Of the five GBBs connected to the most source fields, three concern construction materials, laminates, or composites; the other two concern measurement and medical technologies. Three of the five GBBs connected to the fewest source fields concern molecular sieves or catalysts. Source fields connected to the fewest GBBs mostly concern vehicle or engine components, whereas those connected to the most include heating, drying, lasers, electromechanical technologies, and chromatography (SI Section~\ref{sec:SI_network_structure}).

Fig.~\ref{fig:fig2c} shows how these findings generalize. The figure depicts the adjacency matrix of the bipartite network between GBBs (rows) and source fields (columns). The structure of this matrix is somewhat triangular, a characteristic that is referred to as \emph{nested} in ecology\cite{almeidaneto_consistent_2008} and other fields\cite{mariani_nestedness_2019}. Each GBB connects to between 10 and 61 source fields, while each source field connects to between 4 and 29 GBBs. Thus, some GBBs can be applied to a wide range of source fields, whereas others have a narrower range of observed applications. Likewise, for some source fields we observe many available GBB options, whereas for others we observe few.
Such heterogeneity shows that the breadth of available options for decarbonization differs substantially across domains.

\begin{revision} We conjecture that the abundance of decarbonization options available in high-degree source fields allows industries to more easily circumvent technological bottlenecks. In contrast, a low-degree source field can draw only on a narrow set of observed GBBs. If the associated decarbonization strategies become infeasible or expensive, progress may slow down. Conversely, a high-degree GBB is reusable across many application contexts, while a low-degree GBB is specialized. Improvements in the former may therefore impact a wider set of activities than improvements in the latter.\end{revision}

\subsection*{\textcolor{red}{Predicting diversification of green innovation}}
If GBBs reflect decarbonization capabilities, they should be predictive of how firms innovate. To show this, we use regression analysis to predict how firms diversify into new Y02/04 technology classes. Observations consist of firm$\times$green-patent-class combinations and are limited to those where the firm held no prior patents in the CCMT class in the base period. The dichotomous dependent variable encodes the event in which the firm starts patenting in the green technology class thereafter. In all models, we control for firm and CCMT technology class fixed effects, such that the estimation only uses information on the relative differences among technologies within the same firm. 

\begin{figure}[htp]
\centering
    \setlength{\tabcolsep}{3pt}
    \begin{subfigure}[b]{\linewidth}
        \centering
        \input{tab/tab1}
        \caption{Regression Table}\label{table1}
    \end{subfigure}
    \vfill
    \begin{subfigure}[b]{\linewidth}
        \centering
        \includegraphics[width=\linewidth]{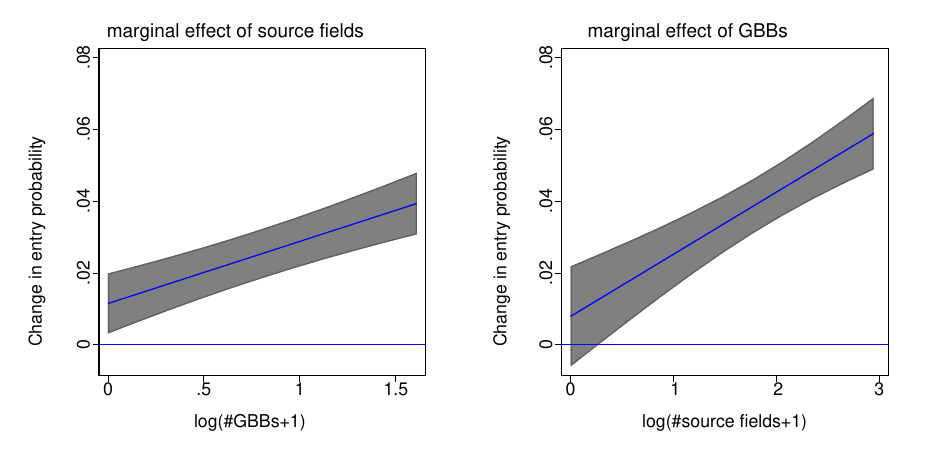}
        \caption{Marginal effect}\label{fig3b}
    \end{subfigure}
\caption{Green innovation by firms.
\textbf{(\subref{table1})}: regression analysis of whether or not a firm patents in a new CCMT  class. Independent variables: $\log(N_{f,src(c),0}+1)$ : \textcolor{red}{logarithm of one plus the number of source fields relevant to the CCMT class in which the firm specializes}; $\log(N_{f,GBB(c),0}+1)$ : \textcolor{red}{logarithm of one plus the number of GBBs relevant to the CCMT class in which the firm specializes}; $\log(N_{f,src(c),0}+1) \times log(N_{f,GBB(c),0}+1)$  interaction term. Note that to avoid $\log(0)$ observations, we augment each count by 1. All models include firm and CCMT fixed effects. \textbf{(\subref{fig3b})}: interaction plots, showing how parameter estimates vary over the range of the moderating variable observed in the dataset. Shaded areas indicate 95\% confidence intervals computed from two-way cluster-robust standard errors (firm and CCMT). All tests are two-sided. \textcolor{red}{The estimates show that prior source-field and GBB portfolios jointly predict subsequent entry.}}
\label{fig:fig3}
\end{figure}

Figure~\ref{fig:fig3} summarizes results.  The estimated coefficients express the change in the probability that a firm enters a specific new CCMT technology class associated with a one log-point increase in the regressor value. The explanatory variables express the technologies in which the firm has patented. For each CCMT technology, we first identify \emph{relevant GBBs}, i.e., GBBs that connect to the CCMT class in the tripartite network of Fig.~\ref{fig:fig1}. Next, we identify source fields that connect to these GBBs as \emph{relevant source fields}. 

Both prior specialization in relevant nongreen source fields (column 1 of the table in Fig.~\ref{fig:fig3}) and in relevant GBBs (column 2) are associated with an increased likelihood that the firm patents in a CCMT class. However, the GBB effect is over three times as strong as the source field effect (column 3). Moreover, both effects reinforce one another (column 4). This is shown graphically in the lower panels of Fig.~\ref{fig:fig3}. The left panel shows how the source field effect changes with the number of relevant GBBs in which the firm specializes. The right panel shows how the GBB effect changes with the number of relevant source fields in which the firm specializes. Prior specialization in relevant source fields is associated with increased green patenting at any level of GBB specialization, whereas specialization in relevant GBBs only matters once the firm is also specialized in relevant source fields. \begin{revision}This interaction is consistent with firms' applying GBBs in their own source fields.
Controlling for how related a green technology is to the firm's patent portfolio, the source–GBB interaction remains positive at 0.01099 (95\% CI [0.00257, 0.01941]), showing that GBBs add information about subsequent green entry beyond ordinary technological relatedness(SI, Section~\ref{sec:SI_entry_relatedness}).\end{revision}

\subsection*{\textcolor{red}{Technological complementarity and alliances}}
When firms lack relevant GBBs, they may still access them through technological alliances with other organizations. To study this, we identify for each firm with at least 500 patent families the most suitable collaboration partner from the perspective of GBBs. That is, we identify the most \emph{complementary} organization (firm or otherwise), i.e., the organization that can offer the largest number of \emph{missing} GBBs: GBBs that are relevant to the firm's source fields but in which it has not developed any expertise itself (see Methods). The greater the number of missing GBBs a potential partner can offer, the better the match.  

\begin{revision}The unit of analysis in the alliance-probability comparison is a focal firm--candidate organization pair. Complementarity is ranked separately within each focal firm's set of 7,189 other eligible organizations, such that a firm's most complementary organization --- the organization supplying the largest number of missing, yet relevant, GBBs --- is ranked 1.\end{revision}

\begin{figure}[htp]
\centering
\begin{subfigure}[b]{0.35\linewidth}
    \centering
    \includegraphics[width=\linewidth]{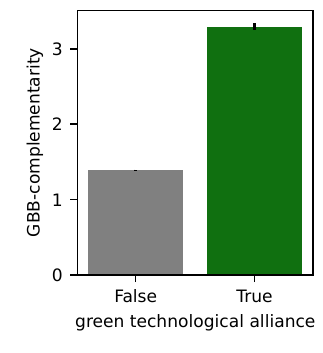}
    \caption{}\label{fig4a}
\end{subfigure}
\hfill
\begin{subfigure}[b]{0.5\linewidth}
    \centering
    \includegraphics[width=\linewidth]{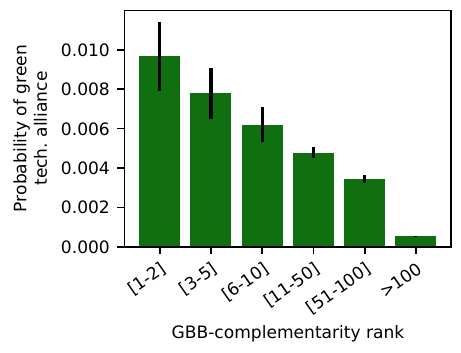}
    \caption{}\label{fig4b}
\end{subfigure}
\vfill
\begin{subfigure}[b]{0.45\linewidth}
    \centering
    \includegraphics[width=\linewidth]{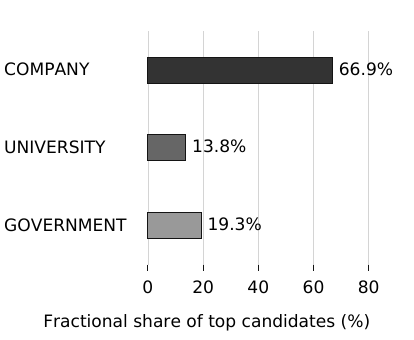}
    \caption{}\label{fig4c}
\end{subfigure}
\hfill
\begin{subfigure}[b]{0.45\linewidth}
    \centering
    \includegraphics[width=\linewidth]{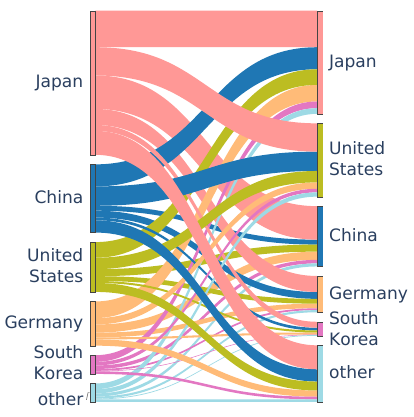}
    \caption{}\label{fig4d}
\end{subfigure}
\caption{Technological alliances. \textbf{(\subref{fig4a})}: Average GBB-based complementarity (measured in \# GBBs) in green technological alliances compared to random pairings of firms and organizations. Bar shows means, and vertical lines are 95\% confidence intervals (alliances: n = 25,525; comparison pairs: n = 42,727,458). \textbf{(\subref{fig4b})}: Probability of observing a green technological alliance between potential partners at different ranks of GBB-based complementarity. Ranks are grouped into bins along the horizontal axis, and vertical lines represent 95\% confidence intervals. \textbf{(\subref{fig4c})} \textcolor{red}{Types of most complementary candidate organizations.} \textcolor{red}{Percentages use fractional counts across 5,947 selected firm--candidate pairs. GOV and NON-PROFIT always co-occur in these pairs and are represented by GOV only. Each pair contributes a total weight of one, divided equally among its remaining organization-type labels.} \textbf{(\subref{fig4d})}: Sankey diagram representing the number of \textcolor{red}{most complementary candidates} located in countries on the right for firms in countries on the left. \textcolor{red}{Observing co-assigned green patents is substantially more common among pairs with greater GBB complementarity.}}
\label{fig:fig4}
\end{figure}

We can observe technological alliances when two organizations take co-ownership of the same patent. In that case, the patent is assigned at issue to two different firms. We focus on alliances in green technologies, such that we observe a technological alliance if two organizations are co-assignees on at least one green patent. \begin{revision}When we observe such co-assigned green patents, we refer to this as an ``alliance'' between the assignees.\end{revision}

Figure~\ref{fig:fig4} shows that \textcolor{red}{observing such alliances is more common among} \begin{revision}more complementary candidate pairs.\end{revision} On average, the complementarity ---measured as the number of missing GBBs a potential partner can offer --- is 3.295 (SE = 0.0230; 95\% CI: 3.250--3.340) among organizations that form a green technological alliance, versus 1.398 (SE = 0.0003; 95\% CI: 1.397--1.398) among organizations without such alliances. Similarly, among all potential organizational pairs, the likelihood of a green technological alliance increases by a factor of 17.8 as we move from the least to the \begin{revision}most complementary pairings. This association also holds in linear probability regressions after accounting for focal-firm and candidate-organization fixed effects (coefficient 0.00033, $p<0.001$; SI Table~\ref{tab:pred_cogreenpat}) and when we remove some pairs that may belong to the same corporation  (SI, section~\ref{sec:APP_observed_samegroupaudit}). \end{revision} \begin{revision}GBBs thus help predict which organizations form alliances to develop green patents together.\end{revision}

\subsection*{\textcolor{red}{Green building blocks as decarbonization capabilities}}

\begin{figure}[htp]
\centering
\begin{subfigure}[b]{\linewidth}
    \centering
    \includegraphics[width=\linewidth]{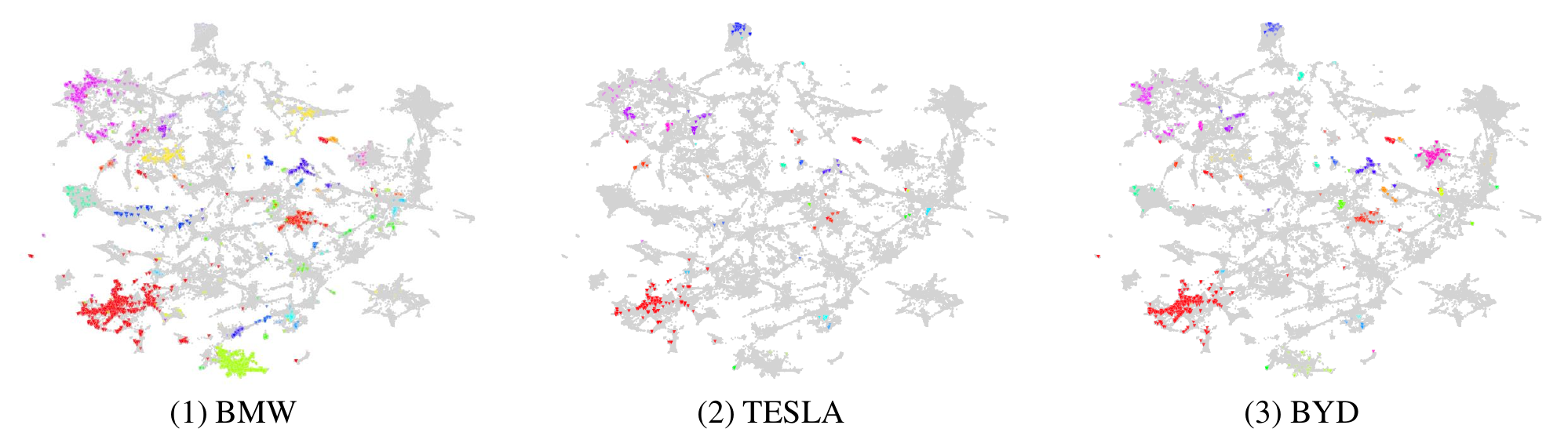}
    \caption{Green building blocks of 3 vehicle manufacturing firms}\label{fig5car}
\end{subfigure}
\vfill
\begin{subfigure}[b]{\linewidth}
    \centering
    \resizebox{\textwidth}{!}{\input{tab/tab_firm}}
    \caption{Top green building blocks of 3 vehicle manufacturing firms by revealed comparative advantage }\label{table2}
\end{subfigure}
\vfill
\begin{subfigure}[b]{\linewidth}
    \centering
    \includegraphics[width=\linewidth]{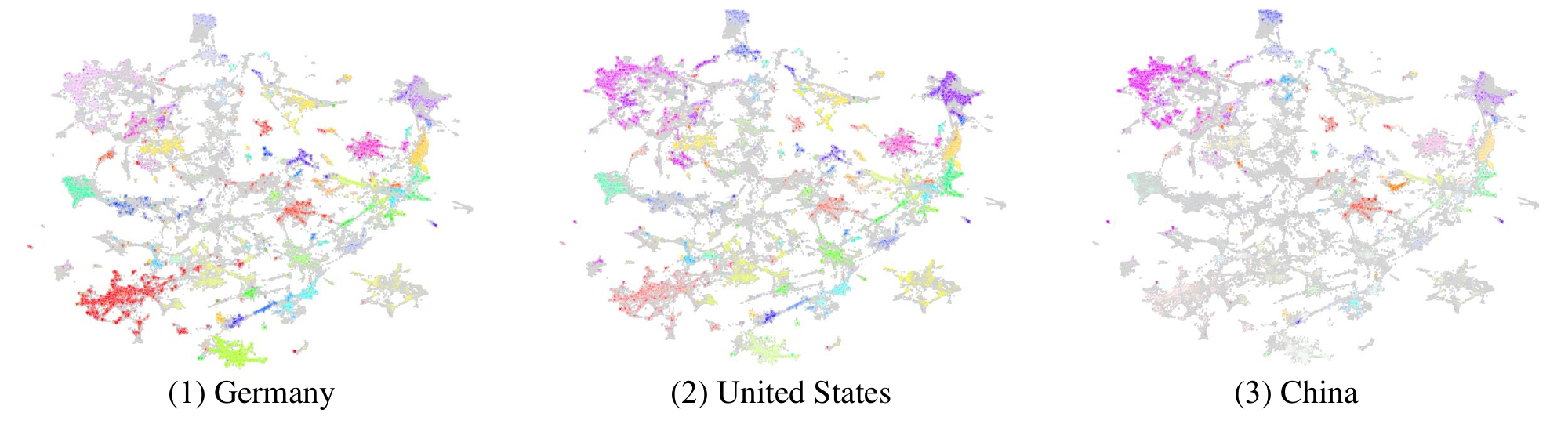}
    \caption{Green building blocks of 3 countries}\label{fig5cntry}
\end{subfigure}
\vfill
\begin{subfigure}[b]{\linewidth}
    \centering
    \resizebox{\textwidth}{!}{\input{tab/tab_cntry}}
    \caption{Top green building blocks of 3 countries by revealed comparative advantage}\label{table3}
\end{subfigure}
\caption{Green capability bases of selected firms and countries. Layout generated by UMAP, colors indicate GBBs and transparency scales vary with GBBs' overrepresentation in the firm or country (based on RCA, see Methods). \textcolor{red}{The figure shows that firms and countries combine distinct, unevenly distributed portfolios of climate-mitigation components.}}
\label{fig:fig5}
\end{figure}

Taken together, these analyses suggest that GBBs capture  relevant climate-change mitigation capabilities. Fig.~\ref{fig:fig5} shows that these capabilities differ substantially across countries and firms. 

The top part of Fig.~\ref{fig:fig5} displays the GBBs of three major car manufacturers ---  BMW, Tesla and BYD --- highlighting  the GBBs in which each car firm is overrepresented, when comparing the share of the firm's patents in this GBB to \textcolor{red}{the GBB share among the 1,422 automotive applicants in the comparison group (see Methods)}. BMW’s strength in green innovation is mainly rooted in optimizing traditional automotive technologies, with GBBs in \emph{advanced engine systems} and \emph{power conversion}, but also in electric vehicle and fuel cell related technologies (e.g., \emph{electrochemical energy storage} and --- hydrogen tanks related --- \emph{gas storage and transfer}). In contrast, Tesla's decarbonization capabilities consist in integrating multiple emerging fields in sustainable technology from \emph{electric motor control} to \emph{solar energy} and \emph{electrical coupling}. Finally, BYD's GBBs (e.g., \emph{material science} and \emph{electrochemical energy storage}) betray the firm's strength in core battery and EV technologies.

The bottom part of Fig.~\ref{fig:fig5} shifts the analysis to the level of countries, highlighting which GBBs are exceptionally well developed in Germany, the US and China. Although all three countries leverage various GBBs, Germany and the US have a greater variety of GBBs at their disposal than China. However, the type of decarbonization capabilities differs markedly. Germany specializes in GBBs associated with its traditionally strong automobile and petrochemical industries (e.g., \emph{advanced gasification}, \emph{automotive mechanism cluster} technologies and  \emph{engine systems}). The most overrepresented GBBs in the US include \emph{hydrocarbon processing}, and technologies related to \emph{medical devices} and \emph{biotechnology processes and compounds}. Finally, China specializes in GBBs related to its rapid expansion in infrastructure and urbanization, such as \emph{advanced concrete materials}, but also to several GBBs in electronics, such as  \emph{advanced solar heat collection},  and \emph{lighting technologies}.

Combined with the links between GBBs and green technologies, such portfolios indicate where each country's existing capabilities are most likely to be useful. For instance, Fig.~\ref{table3} suggests that Germany could look for green applications of its capabilities in gasification, engine systems and steam systems, the US of those in hydrocarbon processing, catalysis and biotechnological processes, and China of those in concrete materials, network management and solar heat collection.

However, this presumes that firms can only access GBBs inside their own countries. If, instead, firms can access GBBs abroad through international alliances, the option space for green innovation widens substantially. In fact, Fig.~\ref{fig4d} shows that the \begin{revision}most complementary candidates\end{revision} often reside outside firms' home countries. For instance, 38.6\% of \begin{revision}the most complementary candidates\end{revision} for US firms reside in Japan and a further 26.4\% in China (see Table~\ref{tab:best_supplier_cntry} of the SI). \begin{revision}This prominence of Japan is in line with its well-documented historical strength in green patenting\cite{dechezlepretre2011invention}.\end{revision} Moreover, the most complementary partners are not always other firms:  \textcolor{red}{Universities account for 13.8\% of most complementary candidates under fractional counting.}
Judged by their GBBs, highly complementary partners are thus often found across national and sectoral borders. This suggests that firms aiming to reduce their carbon footprint may benefit from considering partners beyond their home country and outside the private sector.

\section*{Discussion}
Innovation often consists of new combinations of existing ideas. Applying this insight to the green transition, we have proposed that many climate-change mitigation technologies combine two components: nongreen source technologies  and GBBs that  mitigate their adverse impact on the global climate. GBBs can therewith be interpreted as generalized capabilities that can be used to reduce carbon emissions across a range of nongreen applications. 

At a methodological level, this perspective casts the green transition as a bipartite network in which nongreen source technologies are  connected to the GBBs with which they can be combined. \begin{revision}This offers an empirical meso-level\cite{dopfer_micro-meso-macro_2004} perspective that complements the system-wide representations in IAMs and the application-specific detail of engineering studies.\end{revision}

The network connecting nongreen sources to GBBs turns out to be nested. This has important implications for the anatomy of the green transition, where problems differ in the length of the menu of  available climate-change mitigation pathways and solutions differ in the range of problems they can help resolve. \begin{revision}Our study makes this heterogeneity precise: source fields linked to few GBBs have a comparatively narrow set of observed mitigation options. This points to potential bottlenecks that may require prioritizing the development of additional solutions.\end{revision} Similarly, we are able to distinguish between general-purpose GBBs and GBBs that can only be applied to specific problems. \begin{revision} The network thus identifies where cross-domain synergies may materialize and where they are likely to be sparse.\end{revision}

GBBs may be a useful construct for strategic management in firms engaged in the green transition. First, a firm's prior activity in nongreen source fields and in GBBs predicts which CCMTs it develops. \begin{revision}Our findings suggest that firms apply  GBBs to the (nongreen) source technologies in which they are active.\end{revision} Second, GBBs can help identify suitable partners for technological alliances: we find that firms  disproportionately collaborate with organizations that hold GBBs they themselves need, but lack, to reduce the climate impact of their own source technologies.

\begin{revision}At the country level, GBB portfolios identify technological strengths on which further green innovation can build.\end{revision} For instance, our analysis suggests that Germany's green transition can leverage GBBs related to chemical processes and traditional automobile manufacturing, whereas  the US can build on GBBs in chemistry and  biotechnology. Electronics and Networks constitute the main strength of the Chinese economy. However,  opportunities widen significantly if firms can engage in cross-border technological alliances. \begin{revision}The reason for that is that the most complementary candidates often lie outside the firm's home country.\end{revision} \begin{revision}The GBB framework can help identify such opportunities for international and cross-sectoral knowledge access.\end{revision}

Our analysis has several limitations. First, our pairing of green to nongreen patents based on the functions of the patented technologies focuses on improving existing nongreen technologies. However, the methodology is less suited to analyze systemic change that leads to, for instance, completely new transportation or power generation systems. Furthermore, our reliance on patents emphasizes manufacturing processes, where innovations are more often patented than in new business models or organizational processes. However, such organizational innovations can also offer large gains in efficiency and shift consumption to less carbon intensive activities. Finally, although we have experimented with several different ways to identify GBBs, we expect that further analysis can improve the framework's signal-to-noise ratio.    

\begin{revision}The GBB framework can be further developed with the aim of supporting decisions about green technology development.\end{revision} First,  GBBs can help policymakers quickly obtain an overview of the strengths and weaknesses of their economies in terms of green innovation. Second, \begin{revision}the network complements detailed engineering studies and high-level IAM pathways by identifying recurrent components and their observed domains of application.\end{revision} Third, \begin{revision}source fields connected to few GBBs can be flagged as potential future bottlenecks in the green transition. Similarly, the network can identify broadly reused components to assess their accessibility and scope for diffusion.\end{revision} Moreover, although we have analyzed a single network for the entire period, our framework can also be used to trace how GBBs and their connections to source fields evolve over time. This is a fruitful area for future research, which would support such decision making in a world where technological opportunities keep changing. Finally, \begin{revision}governments can identify groups of firms and research institutes for potential synergies.\end{revision} Companies, in turn, \begin{revision}can use the network to screen partners or customers to assess the relevance of their GBB expertise.\end{revision}

\section*{Methods}

\subsection*{Data}

To construct GBBs, we use 8.4 million patents  with full text information in English between 1976 and 2022, downloaded from PatentsView/USPTO \cite{uspto_patentsview_2024}.  PATSTAT Simple family IDs (DOCDB) are used to de-duplicate patents within the same family, which results in 5.7 million patent families, of which we classify \begin{revision}288,579\end{revision} as CCMT or ``green'' patents. USPTO patent documents include several text fields, including a title, abstract, summary and further details. These fields describe the patented technology in an increasingly detailed manner. 

\begin{revision}Patent texts are taken from PatentsView, while firm portfolios and co-assignment relationships are constructed from PATSTAT's recorded links between applicants and applications. To identify applicants, we do not rely on the raw applicant-name strings, but on PATSTAT's Standardised Name ID (\texttt{psn\_id}). psn\_ids are assigned by PATSTAT to applicants, based on a name-harmonization procedure developed by ECOOM at KU Leuven. Different spellings of an applicant's name that this procedure recognizes share one psn\_id, and we treat them as one organization. Note that this does not necessarily consolidate different legal entities belonging to the same corporate group, leaving some ambiguity about whether two distinct applicants are part of the same firm.\end{revision} 

Patent documents contain technology codes, describing the  technological  fields relevant to the invention. We rely on the CPC (Cooperative Patent Classification) classification developed by USPTO and EPO. This classification includes so-called "Y02/04" classes that identify climate-change mitigation technologies (CCMT). SI Fig.~\ref{fig:pairsample} provides an example of paired CCMT and nongreen patents.

Finally, to count the number of patents -- identified as PATSTAT Simple family IDs (DOCDB) -- assigned to a firm or produced in a country, we rely on the EPO PATSTAT dataset \cite{epo_patstat_2019}, which has better data coverage. \textcolor{red}{We use the country of residence of inventors to characterize countries' patent portfolios.} Because patents are granted with a delay, we exclude the last years of the PATSTAT data, focusing on the \textcolor{red}{1995--2004} and \textcolor{red}{2005--2014} periods in our regression analyses.

\subsection*{Pairing CCMT and nongreen patents with similar functions}\label{method:paring}

Comparing CCMT patents to nongreen patents with similar  functions is complicated because patent applications do not list a section that explicitly describes the functionality of a patented technology. For instance, the CPC codes on a patent describe  implementation details (``how the functionality is achieved'') not the functionality itself. Therefore, we develop a method to assess the similarity of patents in terms of the functionality of the invention they describe.

Patent texts often do describe their invention's functionality in free text format, such as ``this invention is to solve the problem of X.'' Or more implicitly, the function of a patent can be inferred from the description of  application scenarios. 

To analyze such textual information, we use text embeddings to convert the text on a patent in a dense, high dimensional vector. Compared to similarities in terms of keyword or word co-occurrences, such embeddings take the semantic meaning of a text into consideration. In particular, we use the Python package \emph{FSE}'s CBOW model on the PARANMT-300 pre-trained embedding to convert four different text fields on each patent into high-dimensional vectors: (1) the title and abstract field (``abstract''), (2) the summary field, which summarizes the invention (``summary''), the claims field, which describe legal claims that delineate the novelty of a patent (``claim'') and (4) the detailed description of the patent (``detail''). Next, we calculate the similarity of a CCMT patent to all nongreen patents in the database in terms of the \textcolor{red}{cosine similarity} of the embeddings of these four different text blocks. To verify that these \textcolor{red}{cosine similarities} indeed rank pairs of patents in terms of how similar they are in terms of an invention's functionality, we compare them to rankings provided by human experts (see SI section \ref{sec:SI_vali_human} for details). Overall, the  embedding for the abstract text turned out to align best with the assessments by human raters. We therefore rely on these embeddings when we pair CCMT patents to nongreen counterparts with highly similar functions.

\subsection*{Extracting GBBs from paired patents}

We use these CCMT--nongreen patent pairs to extract CPC codes that are frequently found on the CCMT patent but not on its matched nongreen patent (the ``extra'' CPC codes). However, in this process, some complications arise that need to be dealt with. 

First, ideally we use the most detailed  CPC technology classification codes available to extract fine-grained ``extra'' CPC classes. The problem is that at the most granular level of the classification system, even highly similar patents typically list very few CPC codes that are exactly the same. Instead, these patents would list codes that are very similar. For example, patents related to battery technology could list either CPC code H01M 6/14 (Battery Cells with non-aqueous electrolyte), or H01M 6/18 (Battery Cells with solid electrolyte). Even when  both codes would apply equally well, patents typically list only one of these alternatives. Second, when pairing patents to each other, it is impossible to avoid measurement errors or noise related to the quality and size of the sample from which to select plausible matches,  mistakes in the original code assignment, and suitability of the metric used to assess similarity. 

The problem arising from working with granular technology classifications is very similar to the problem of dealing with words in Natural Language Processing (NLP), where two different words might express similar meaning. The NLP literature has resolved this by using algorithms such as  Word2Vec, which generate a vector representation for each word in a high-dimensional continuous space. This allows finding words that are semantically similar (i.e., synonyms) by calculating the distance between the embedding vectors of these words.

In analogy, we train a CPC2vec model using the Word2vec algorithm in Gensim, which creates a 50-dimensional embedding vector for each CPC code. The training was based on the CPC codes in the PATSTAT 2019 database, using a skipgram model that uses a  focal CPC code to predict other CPC codes on the same patent. Consequently, codes with a similar usage pattern on patents will be close to each other in the 50 dimensional space.

A visual representation of this space can be produced by reducing the 50-dimensional space to a 2-dimensional space using the UMAP algorithm. The result is the network depicted in Fig. \ref{fig:fig1}, where the markers depict individual CPC codes. The same  visual representation is also used in Fig.~\ref{fig:fig5} to highlight GBBs of firms and countries.

\begin{revision}
To reduce noise in the extraction of ``extra'' CPC codes, we compare each code $c$ on a CCMT patent $p$ with each code in the set of codes $S_b$ on an individual matched nongreen patent $b$. With cosine similarity $s^{\mathrm{CPC}}$ and cosine distance $d^{\mathrm{CPC}}=1-s^{\mathrm{CPC}}$, the score is the distance from the closest counterpart:
\[
L_{cpb}=1-\max_{c'\in S_b}s^{\mathrm{CPC}}_{c,c'}
       =\min_{c'\in S_b}d^{\mathrm{CPC}}_{c,c'}.
\]
The CPC2vec vectors are normalized before these similarities are calculated. We select a code within a patent pair when its score exceeds $Q_{.75}+1.5(Q_{.75}-Q_{.25})$, where the quartiles are computed across the CCMT patent's valid non-Y CPC codes. This procedure is repeated for each match. Results are then aggregated across matches.

Across the matched patent pairs, this procedure identifies 37,801 distinct CPC codes as additions and 75,636 as source technologies. \end{revision}

To identify GBBs, we group closely related added technologies based on their similarity in the CPC2vec embedding space. To do so, we use the density-based HDBSCAN algorithm. Following standard practice, we first reduce the dimensionality of the CPC2vec vectors from the original 50 dimensions to 5 dimensions using UMAP. This  reduces the computational burden and the noise in the dataset. \textcolor{red}{We set the minimum cluster size to 100 CPC-code points to extract larger meaningful groups.} This clusters technology codes into 82 GBB technology clusters (``GBBs'') and 193 source field clusters (``sources''). To reference individual GBBs and sources in the paper, we label them by prompting  a Large Language Model (GPT 4o) to summarize \textcolor{red}{up to 30 CPC descriptions per cluster. For GBBs, we rank these descriptions using $q_c$, which gives greater weight to codes identified as additions in more textually similar patent pairs (see SI, Section~\ref{sec:SI_identify_gbb}). The prompt requests labels of fewer than five words and descriptions of fewer than twenty words.} Detailed prompts are included in the SI.

\subsubsection*{Evaluating the effects of GBB in firm innovation}

We describe a firm's CCMT patent portfolio by looking at all patents assigned to the firm. To describe the technology of a patent, we use CPC codes at the most disaggregated level. Next, we split the data into two distinct sample periods: \textcolor{red}{1995--2004} and \textcolor{red}{2005--2014}. We then aggregate the patents in each period to CPC-firm cells, counting how many patents a firm had in each CPC code. 

\textcolor{red}{In the regression analysis of Fig.~\ref{fig:fig3}, the targets are 49 Y02/Y04 main groups; we focus on targets where the firm has no patents in the first period.} Next, we analyze how the likelihood that firms enter (i.e., start patenting in) the CPC code in the second period. We limit this analysis to Y02/Y04 codes.  To do so, we estimate the following Ordinary Least Squares (OLS) model:

\begin{equation} \label{eq:entry}
\begin{aligned}
    1(N_{f,c,1}>0)  = & \beta_{s}log(N_{f,src(c),0}+1) \\
    +&\beta_{g}log(N_{f,GBB(c),0}+1) \\
    +&\beta_{s \times g}log(N_{f,src(c),0}+1)log(N_{f,GBB(c),0}+1) \\
+&\delta_{f} + \gamma_c + \varepsilon_{fc}, 
\end{aligned}
\end{equation}
where we limit the sample to observations where $1(N_{f,c,0}=0)$ and
\begin{itemize}
    \item $1(.)$ an indicator function that evaluates to 1 if its argument is true and 0 otherwise;
    \item $N_{f,c,0}$ the number of patents by firm $f$ in CCMT class $c$ in the base period of \textcolor{red}{1995--2004} 
    \item $N_{f,src(c),0}$ \textcolor{red}{the number of source fields relevant to class $c$ in which firm $f$ has a  baseline specialization}
    \item $N_{f,GBB(c),0}$ \textcolor{red}{the number of GBBs relevant to class $c$ in which firm $f$ has a baseline specialization}
    \item $\delta_{f}$ a firm fixed effect
    \item $\gamma_c$ a CCMT class fixed effect
    \item $\varepsilon$ a disturbance term. 
    
\end{itemize}

\begin{revision}The source and GBB counts use $RCA\geq1$ to identify baseline specialization. In section~\ref{sec:SI_entry_relatedness}, we also control for technological relatedness.\end{revision}

The unit of analysis is the firm $\times$ CCMT-class cell; the estimation sample contains N = 600,571 such cells. All hypothesis tests are two-sided. Because observations of the same firm or same CCMT class may be correlated, we report cluster-robust standard errors, two-way clustered at the firm and CCMT levels. 

\subsubsection*{Highlighting specialized GBBs in firms and countries}
The ``overrepresented'' GBBs of firms and countries in Fig.~\ref{fig:fig5} are highlighted using opacity by their Revealed Comparative Advantage (RCA)  score:

\begin{equation*}
RCA_{ci}=\frac{N_{ci}/N_{c.}}{N_{.i}/N_{..}},    
\end{equation*}
\textcolor{red}{where $N_{ci}$ is the patent count of firm (or country) $c$ in GBB $i$, $N_{c.}$ its total count, $N_{.i}$ the count in GBB $i$ across the comparison group, and $N_{..}$ the grand total for that group.}

For car firms, we used the 1,422 company assignees that have at least 80 distinct patent families with the CPC code B60, F02 and B62D as the comparison group. The comparison group of countries include all countries in PATSTAT.

\subsubsection*{Identifying collaboration partners}
In our analysis of strategic alliances, we restrict the sample to 7,190 organizations with at least 500 patent families, among which 5,947 are firms. Next, we focus on firms as the main unit of analysis. We then ask which other organization (firm or otherwise) has the largest number of GBBs that are relevant to the firm's nongreen source technologies, but are not possessed by the firm itself, among the 42,752,983 pairs of firms with other organizations. \begin{revision}We say that a firm \emph{demands} a GBB if it is specialized in the source fields linked to that GBB. An organization  \emph{possesses} a GBB if it is specialized in the GBB itself. In both cases, specialization requires that the firm or organization has more than ten patent families and $RCA\geq1$. Complementarity is the number of GBBs that the candidate possesses, but that the focal firm does not possess but does demand. Details are given in Section~\ref{sec:SI_alliance_complementarity} of the SI.\end{revision} Next, for each firm, we rank all organizations in descending order of this count. In the case of ties, we use information on the total volume of patents the organization holds in the missing GBBs to break them.

Determining the national origins of firms and organization is complicated by the fact that many companies incorporate in countries with favorable tax regimes. Therefore, we determine for each firm and organization the country in which most of its inventors reside, using not the location reported for the firm or organization, but the location of residence of its inventors. This means that the ``nationality'' of a company from the perspective of the current study is the country where it concentrates most of its research and development activity. 

\begin{revision}We say that we observe a technological alliance between a focal firm and a candidate organization when the two share at least one green patent family. Figure~\ref{fig:fig4} compares the complementarity of pairs with and without such an alliance and shows how the frequency of alliances varies with complementarity rank. Portfolios and alliances are both measured using all patent families in our data. We also estimate linear probability models that regress the presence of an alliance on $\log(1+K_{fo})$, where $K_{fo}$ is the pair's complementarity. This allows us to add the following control variables: an indicator for whether both organizations are located in the same country, the patent volumes of both organizations, and fixed effects for the focal firm and the candidate organization. Standard errors are two-way clustered by focal firm and candidate organization. As a robustness check, we repeat the fixed-effects model without cross-country pairs of companies that a large language model (DeepSeek) flagged as possibly belonging to the same corporate group. Details are provided in Sections~\ref{sec:APP_allianceregression}--\ref{sec:APP_observed_samegroupaudit} of the SI.\end{revision}

\subsection*{Availability of data and code} PatentsView data are publicly available. PATSTAT data require a license. The code for all data processing steps, analyses and figures, including the sensitivity analyses in the SI, is available at \url{https://github.com/complexly/gbb_code} and archived at Zenodo (DOI: 10.5281/zenodo.16552714).

\subsection*{Use of Generative AI} \textcolor{red}{Generative AI was used for proposing cluster names (GPT-4o), language editing (ChatGPT and Claude), functional-similarity judgments (DeepSeek-V4-Flash and GLM-5.3), and name-based corporate-group screening (DeepSeek-V4-Flash). The SI describes how these judgments were obtained. The analysis and coding for all main results were carried out by the authors. The coding of some sensitivity and robustness checks was supported by OpenAI Codex. All manuscript text and interpretations were reviewed by the authors.}

\section*{Acknowledgments}
F.N.~gratefully acknowledges financial support from the Austrian Research Agency (FFG), project \#873927 (ESSENCSE). Y.L.~gratefully acknowledges financial support from the National Natural Science Foundation of China (grant no. 52560029). We thank the research assistants Pengfei Hao and Yuqi Shi for assessing the validity of patent pairs. The authors also thank the Growth Lab at Harvard University for its support and Muhammed Yildirim and Ricardo Hausmann for valuable comments.

\bibliographystyle{abbrv}
\bibliography{references}

\newpage
\appendix
\renewcommand{\thesection}{S\arabic{section}}
\setcounter{figure}{0}
\setcounter{table}{0}
\renewcommand\thefigure{S\arabic{figure}}
\renewcommand\thetable{S\arabic{table}}

\section*{Supporting Information}
\setcounter{page}{1}

\section{CCMT patents list a greater number of technology codes}

In the main text, we assert that CCMT patents, on average, list more CPC codes than nongreen patents. Here, we examine this difference at the population level (i.e., between all CCMT and nongreen patents).

To do so, we analyze USPTO patents, accessed through PatentsView, and their CPC classifications. The CPC system introduced ``Y'' categories to supplement the original categories (designated by letters ``A''-``H''). Inventions that either directly or indirectly contribute to reducing greenhouse gas emissions or actively enhance carbon sinks are marked with codes that start with ``Y02/Y04''. Hereafter, we refer to such patents as \emph{CCMT} or \emph{green} patents, while we call patents without  ``Y02/Y04'' codes  \emph{nongreen patents}.

\begin{figure}[htp]
\centering
\includegraphics[width=0.5\textwidth]{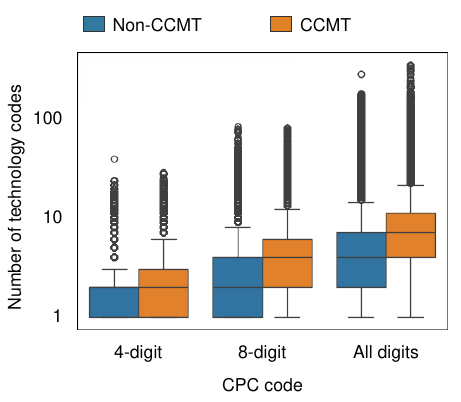}
\caption{More CPC codes in CCMT than nongreen patents}
\label{fig:morecodes}
\end{figure}

On each patent, we count the number of listed CPC codes, excluding CCMT related ``Y'' codes. We interpret this number as an, admittedly imperfect, proxy for the number of technology fields that the invention combines. Figure \ref{fig:morecodes} shows that, on average, CCMT patents list more technology classes than nongreen patents, regardless of the level of aggregation at which we consider  these codes.

Green technologies therefore seem to combine a wider variety of technological fields than nongreen ones, which we interpret as a sign that such technologies are more ``complex.'' However, this interpretation warrants caution, as the observed pattern might simply reflect that green technologies are inherently situated in more complex domains (e.g. machinery or electronics) that require the integration know-how from multiple fields to achieve their intended functionality.

This raises the question of whether green patents list more technology codes than nongreen patents, conditioning on the function their inventions aim to perform. In section~\ref{sec:SI_matching}  below, we present a methodological solution for conducting such controlled comparisons by constructing pairs of green and nongreen patents with very similar functionality. Table~\ref{tab:paired_diff} shows that also within such matched pairs, the green patents list a greater number (on average, 1.77 more) of technology codes than very similar nongreen patents. We view the fields that these ``additional'' technology codes represent as of particular interest.

\section{Pairing CCMT and nongreen patents with similar function}\label{sec:SI_matching}

We analyze 8.4 million USPTO patents, accessed through PatentsView that were granted between  1976-2022. For each patent, PatentsView provides the full English text of the original application. We deduplicate these patents, using the EPO PATSTAT simple family identifiers (DOCDB), which merge all patent documents that refer to the same invention. Furthermore, for all these patents, we use the CPC technology classification codes they list as a succinct and uniform way to characterize the technological know-how involved in the invention.

Unfortunately, there is no straightforward metric in PatentsView that describes the function that a patented technology is supposed to perform.  Instead, the existing classification of CPC codes mostly relate to implementation aspects, i.e., \emph{how} an invention achieves its functionality, not \emph{what} this functionality is in the first place. Therefore, we need to develop a methodology that focuses on the similarity among patents in terms of this functionality as opposed to their technological features. 

\subsection{Approximating functional similarity by text similarity}\label{sec:SI_text_similarity}

Inventors typically describe what their patent can be used for (``its functions'') in natural language -- either explicitly (``this invention aims to solve the problem of ...'') or implicitly by describing application scenarios. While contemporary large language models (GPT/Claude/Gemini) can interpret such text directly, none of these advanced LLMs were available when we initiated this research in 2022. Moreover, using such models to evaluate all pairwise comparisons of millions of patents remains prohibitively expensive and slow. 

Instead, we rely on text embeddings to capture potential functional information of patents in high-dimensional, dense vectors in Euclidean space. Compared to keyword-based methods, word embeddings incorporate semantic meaning, enabling more effective retrieval of functionally similar patents. To construct these embedding vectors, we use the Python package \emph{fse}  with a Continuous Bag of Words (CBoW) model on the PARANMT-300 pre-trained embedding.

For each patent for which PatentsView provides the complete application text, we generate embeddings from four distinct text fields:

\begin{itemize}
    \item Title and abstract (``abstract'')
    \item Summary of invention (``summary'')
    \item Claims (``claims'')
    \item Detailed description (``detail'')
\end{itemize}

This yields four 300-dimensional vectors per patent. Next, for each CCMT patent, we identify the 10 closest nongreen patents based on the cosine similarity of their embedding vectors for each of the four fields. To avoid brute-force computations when dealing with millions of candidates, we use the Hierarchical Navigable Small World (HNSW) algorithm, implemented in the Python package \textcolor{red}{\emph{nmslib}}, for efficient nearest-neighbor retrieval. \textcolor{red}{The routine returns cosine distances, which we convert to similarities as $s=1-d$ for weighting and reporting.} This approach can generate between 10 and 40 candidate nongreen patents, depending on the overlap in matches suggested by the four different text fields.

\subsection{Validation by human raters and large language models} \label{sec:SI_vali_human}

The identification of GBBs crucially relies on being able to match CCMT patents to nongreen patents with comparable functions. To validate the functional similarity scores we use to accomplish this, we compare them to the assessments by two human raters with  expertise in engineering fields. In particular, we asked two graduates of a top Chinese university with a bachelor degree in environmental engineering and sufficient training to understand the meaning of technological terms and their relation to climate change mitigation technologies to assess the similarity between green and nongreen patents.

To do so, we adhere to the following protocol:

\begin{enumerate}
    \item Draw a random CCMT patent, $A$.
    \item Sort all 10-40 candidates of matched nongreen patents by the sum of the cosine similarity of the texts in the abstract, claims, summary and detail sections to arrive at a single similarity score.
    \item Draw two nongreen patents:
    \begin{enumerate}
        \item Draw one patent that is highly similar to $A$, drawing this patent randomly from the top 10\% of $A$'s closest matches.
        \item Draw another patent by first choosing with equal probability among the low-, middle-, and high-similarity tercile groups represented among $A$'s candidates, then randomly drawing a patent from the selected group.
        \item Shuffle the order of the two added nongreen patents and call them $B_1$ and $B_2$
    \end{enumerate}
    \item Repeat steps 1-3 until 200 samples are generated.
\end{enumerate}
Next, we ask each human rater the following question:
\begin{quote}
``Which of patents $B_1$ and $B_2$ is closer to the problem solved by patent $A$, or has a more similar application scenario?''    
\end{quote}
The raters could choose one of three answers: $B_1$, $B_2$ or \emph{not sure}.

\begin{table}[htp]
    \centering
    \caption{Comparison of Agreement with Human Raters}
    \label{tab:combined_agreement}
    \begin{subtable}[b]{\textwidth}
        \centering
        \caption{Average agreement with human raters}
        \label{tab:average_agreement}
        \small
        \begin{tabular}{lccccc}
            \toprule
            Tercile & Inter-rater & Abstract & Summary & Claims & Detail \\
            \midrule
            Same tercile & 0.746 & 0.575 & 0.604 & 0.515 & 0.575 \\
            1 tercile diff & 0.781 & 0.685 & 0.616 & 0.534 & 0.575 \\
            2 tercile diff & 0.854 & 0.659 & 0.732 & 0.488 & 0.659 \\
            All & 0.785 & 0.638 & 0.638 & 0.517 & 0.594 \\
            \bottomrule
        \end{tabular}
    \end{subtable}
    \vfill
    \begin{subtable}[b]{\textwidth}
        \centering
        \caption{Normalized by inter-rater score}
        \label{tab:normalized_agreement}
        \small
        \begin{tabular}{lcccc}
            \toprule
            Tercile & Abstract & Summary & Claims & Detail\\
            \midrule
            Same tercile & 0.770 & 0.810 & 0.690 & 0.770 \\
            1 tercile diff & 0.877 & 0.789 & 0.684 & 0.737\\
            2 tercile diff & 0.771 & 0.857 & 0.571 & 0.771 \\
            All & 0.813 & 0.813 & 0.658 & 0.757\\
            \bottomrule
        \end{tabular}
    \end{subtable}
\end{table}

Results are shown in Table \ref{tab:combined_agreement}. \textcolor{red}{Among the 181 of 200 comparisons in which both raters selected a candidate, they choose the same nongreen patent ($B_1$ or $B_2$) as most similar in function to $A$ in 142 cases (78.5\%).} Moreover, this inter-rater agreement is especially high when the cosine similarity scores of $B_1$ and $B_2$ are very different.

Comparing the choices of our human raters to those picked by our algorithm, we find that the texts in the abstract and in the summary yield similarity scores that are most in agreement with the human raters. Because not all patents have a summary field, we choose the abstract as the basis for matching green to nongreen patents. That is, for each CCMT patent, we choose the 10 closest nongreen matched patents. Furthermore, to improve this match, we  require that the matched patents have at least one 4-digit CPC code in common with the CCMT patent and a cosine similarity to the CCMT patent of between 0.5 and 0.99. This avoids choosing irrelevant matches or matches that are essentially the same texts.

\begin{revision}

To extend the analysis in a cost-effective way, we first replicate and then expand the evaluation of the similarity ranks using large language models (LLMs). For the first step, we use the 200 comparisons that have been rated by human raters. We use these cases to assess how well two LLMs, DeepSeek-V4-Flash and GLM-5.3, both run at temperature zero, perform as judges. 

Our procedure is as follows. Each LLM first separately describes the technical problem, intended function, and application context of the focal patent and both candidates. The LLMs see only the patent titles and abstracts and the evaluation instructions, not how the candidates were selected or how the humans rated them. Based on these descriptions they then choose the candidate that is functionally closest to the CCMT patent.

The LLMs agree with individual human raters in approximately 71\% of comparisons for which both the LLM and the human rater pick a match (i.e., do not pick the option ``not sure''). This is relatively close to the inter-rater agreement for human raters of 78.5\%, which arguably describes a natural benchmark. Moreover, the LLMs perform better where human raters agree: in the 142 comparisons in which both human raters selected the same candidate, each LLM selects that candidate in 112 cases (78.9\%). The prompt asks each model to report its confidence on a scale from 1 to 5. Performance is better still when the LLMs report high confidence. Among the 36 comparisons in which DeepSeek reported confidence 5 and both human raters agreed, DeepSeek selected the same candidate in 34 cases (94.4\%).

For the second step, we add 800 new comparisons keeping the original sampling design. For each green patent, we rank the candidate nongreen patents by the sum of their text-similarity scores across four sections: title and abstract, summary, claims, and detailed description. As before, we then draw one candidate, C1, from the top 10\% in terms of this score, and another, C2, randomly from the lower, middle, or upper third in the score distribution, where each tercile is picked with equal probability. We then ensure that the order in which C1 and C2 are presented to the LLMs is exactly balanced. 

Before going into the results, we note that the two LLMs give consistent ratings: across all 1,000 comparisons, the two LLMs agree in 894 of the 978 cases in which neither returns ``not sure'' (91.4\%; 95\% CI: 89.5--93.0\%; Cohen's $\kappa=0.828$). Where human rater scores are available, LLMs agree more often with one another when the human raters do so as well (93.0\%) than when they don't (81.6\%). 

Table~\ref{tab:llm_pair_validation} reports the LLMs' agreement with the human raters and with the text-based rankings. The LLM judges support the validity of our patent matching approach. Disregarding ``not sure'' responses, DeepSeek and GLM select the candidate with the higher summed text-similarity score in 71.0\% and 72.1\% of comparisons. These rates are well above the 50\% that we would expect had the selection been random. Giving the LLM the full patent text instead of title and abstract does not improve its judgments. In fact, DeepSeek does worse, agreeing with human raters less often (64.8\% instead of 71.3\%), which is why we do not pursue this option further.

\begin{table}[htb]
\centering
\caption{Independent validation of the FSE-based functional-similarity ranking. DS denotes DeepSeek-V4-Flash, GLM denotes GLM-5.3. Panel A uses the original 200 human-rated tasks; the row for agreement with both human raters uses the 142 tasks in which they selected the same candidate. Panel B uses 800 new tasks. ``Not sure'' choices are excluded from each judge's denominator, as in the human-validation protocol. Panel A compares LLM choices with human ratings; Panel B compares their choices with text-based rankings in the additional sample.}
\label{tab:llm_pair_validation}
\small
\begin{tabular}{lcc}
\toprule
 & DS & GLM \\
\midrule
\multicolumn{3}{l}{\emph{Panel A: agreement with human ratings}} \\
Human rater 1 & 0.698 & 0.714 \\
Human rater 2 & 0.727 & 0.711 \\
Average across individual raters & 0.713 & 0.713 \\
Agreement with both human raters & 0.789 & 0.789 \\
\addlinespace
\multicolumn{3}{l}{\emph{Panel B: agreement with the text-based rankings}} \\
Higher composite text-similarity score & 0.710 & 0.721 \\
Higher abstract similarity & 0.661 & 0.668 \\
\bottomrule
\end{tabular}
\end{table}

\end{revision}

\subsection{Example of CCMT and nongreen patent with similar function}

To illustrate the type of pairings we arrive at, Figure~\ref{fig:pairsample} shows an example of a pair of CCMT and nongreen patents that describe inventions that perform very similar functions in iron production. The cosine similarity between the two patents is 0.765.

\begin{figure}[htp]
\centering
\begin{mdframed}[linecolor=black, linewidth=1pt]
    \begin{minipage}[t]{0.48\textwidth}
        \textbf{US5989308}\\
        \textbf{Title}: Plant and process for the production of pig iron and/or sponge iron\\
        \textbf{Assignee}: Primetals Technologies Austria GmbH\\
        \textbf{Year}: 1994\\
        \textbf{CPC codes}: C21B13/0033, C21B13/14, C21B13/002, C21B2100/282, C21B2100/44, C21B2100/66, Y02P10/122, Y02P10/134\\
        \textbf{Abstract}: A plant for the production of pig iron and/or sponge iron includes a direct-reduction shaft furnace for lumpy iron ore, a melter gasifier, a feed duct for a reducing gas connecting the melter gasifier with the shaft furnace, a conveying duct for the reduction product formed in the shaft furnace connecting the shaft furnace with the melter gasifier, a top-gas discharge duct departing from the shaft furnace, feed ducts for oxygen-containing gases and carbon carriers running into the melter gasifier and a tap for pig iron and slag provided at the melting vessel. In order to be able to process not only lumpy ore, but also fine ore within a wide variation range with regard to quantity in a manner optimized in terms of energy and product, the plant includes at least one fluidized bed reactor for receiving fine ore, a reducing-gas feed duct leading to the fluidized bed reactor, an offgas discharge duct departing from the fluidized bed reactor and a discharge means provided for the reduction product formed in the fluidized bed reactor, wherein the top-gas discharge duct of the shaft furnace and the offgas discharge duct of the fluidized bed reactor run into a purifier and subsequently into a heat exchanger from which the reducing-gas feed duct of the fluidized bed reactor departs.

    \end{minipage}
    \hfill
    \begin{minipage}[t]{0.48\textwidth}
        \textbf{US5226951}\\
        \textbf{Title}: Method of starting a plant for the production of pig iron or steel pre-material as well as arrangement for carrying out the method\\
        \textbf{Assignee}: Deutsche Voest Alpine Industrieanlagenbau GmbH\\
        \textbf{Year}: 1991\\
        \textbf{CPC codes}: C21B13/002, C21B13/0073, C21B13/023, C21B13/14, F27D17/10, C21B2100/66\\
        \textbf{Abstract}: There is disclosed a method of starting a plant for the production of pig iron or steel pre-material including a direct-reduction shaft furnace and a meltdown gasifier. At first the still empty meltdown gasifier is heated up by aid of a combustible gas and the smoke gases forming are introduced into the still empty direct-reduction shaft furnace. Coke or a degassed coal product is charged into the direct-reduction shaft furnace and the smoke gases introduced into the direct-reduction shaft furnace are passed through the coke or the degassed coal product by releasing their sensible heat. The coke or the degassed coal product thereby is heated to ignition temperature and is charged into the meltdown gasifier in the hot state, catching fire upon the injection of an oxygen-containing gas or of oxygen. A further coal or coke bed serving for gasification is charged on the ignited bed of coke or degassed coal product and the charging substances are charged into the direct-reduction shaft furnace.\\
    \end{minipage}
\end{mdframed}
\caption{Example of paired green and nongreen patent}
\label{fig:pairsample}
\end{figure}

The example shows that the titles and abstracts share many common words across both patents. The shared CPC codes C21B13/002 \emph{Reduction of iron ores by passing through a heated column of carbon}, C21B13/14 \emph{Multi-stage processes processes carried out in different vessels or furnaces}, C21B2100/66 \emph{Heat exchange} implies the source field. The code C21B2100/282 \emph{Increasing the gas reduction potential of recycled exhaust gases by separation of carbon dioxide} exist only in the CCMT patent, which seem critical to reduce the invention's carbon footprint. However, the example also shows that ``extra'' technology codes can be noisy: it is unclear what the relevance of the code C21B13/0033 \emph{In fluidised bed furnaces or apparatus containing a dispersion of the material} is for the green transition.

\subsection{Properties of paired patents}

Table~\ref{tab:paired_diff} shows that there are systematic differences between CCMT patents and their functionally similar matched counterparts. First, the CCMT patents tend to be more recent, with a later application date. Second, the CCMT patents list a larger number of technology codes. Third, the CCMT patents are filed by slightly larger teams of inventors. This suggests that CCMT technologies are newer and more complex than  nongreen patents whose inventions perform similar functions.

\begin{table}[htb]
\centering
\caption{Means of the difference in selected properties between CCMT and their matched nongreen  patents}
\label{tab:paired_diff}
\small
\begin{tabular}{@{}lc@{}}
\toprule
 & Mean Difference \\ 
 & (standard errors in parentheses) \\
\midrule
application year & 3.10 \\
                 & (0.006) \\
\# technology codes & 1.77 \\
                    & (0.004) \\
inventor team size & 0.13 \\
                     & (0.001) \\
\bottomrule
\end{tabular}
\end{table}

\section{Extract GBBs}\label{sec:SI_GBB_construction}

\subsection{CPC2vec, a continuous representation of discrete technology}

In the main text, we define GBBs as sets of ``added'' technology codes that are typically found on green patents, but not on their nongreen counterparts. In practice, technology codes are assigned to patents with a certain amount of noise. Moreover, often, there are several similar technology codes that would be adequate to describe an invention, but patents typically list only one of such close alternatives.

This challenge of working with granular technology classifications parallels issues in Natural Language Processing (NLP), where distinct terms can convey similar meanings. The Word2Vec algorithm offers a solution by creating vector representations of words in a high-dimensional continuous space, using vector similarity to quantify semantic relatedness.

In our work, we develop a CPC2vec model using the Word2vec algorithm implemented in Gensim. This model generates 50-dimensional embedding vectors for each CPC code. The training corpus comprises all CPC codes from the PATSTAT 2019 database, employing a skipgram architecture that uses each focal CPC code to predict other codes appearing within the same patent. This approach ensures that codes with similar usage patterns across patents are positioned close to each other in the 50-dimensional vector space. Note that proximity in this space reflects how codes are used together on patents. It may therefore capture not only how related technologies are, but also conventions in how patent offices and examiners apply the classification \cite{alstott_mapping_2017}.

\begin{figure}[htb]
\centering
\includegraphics[width=0.7\textwidth]{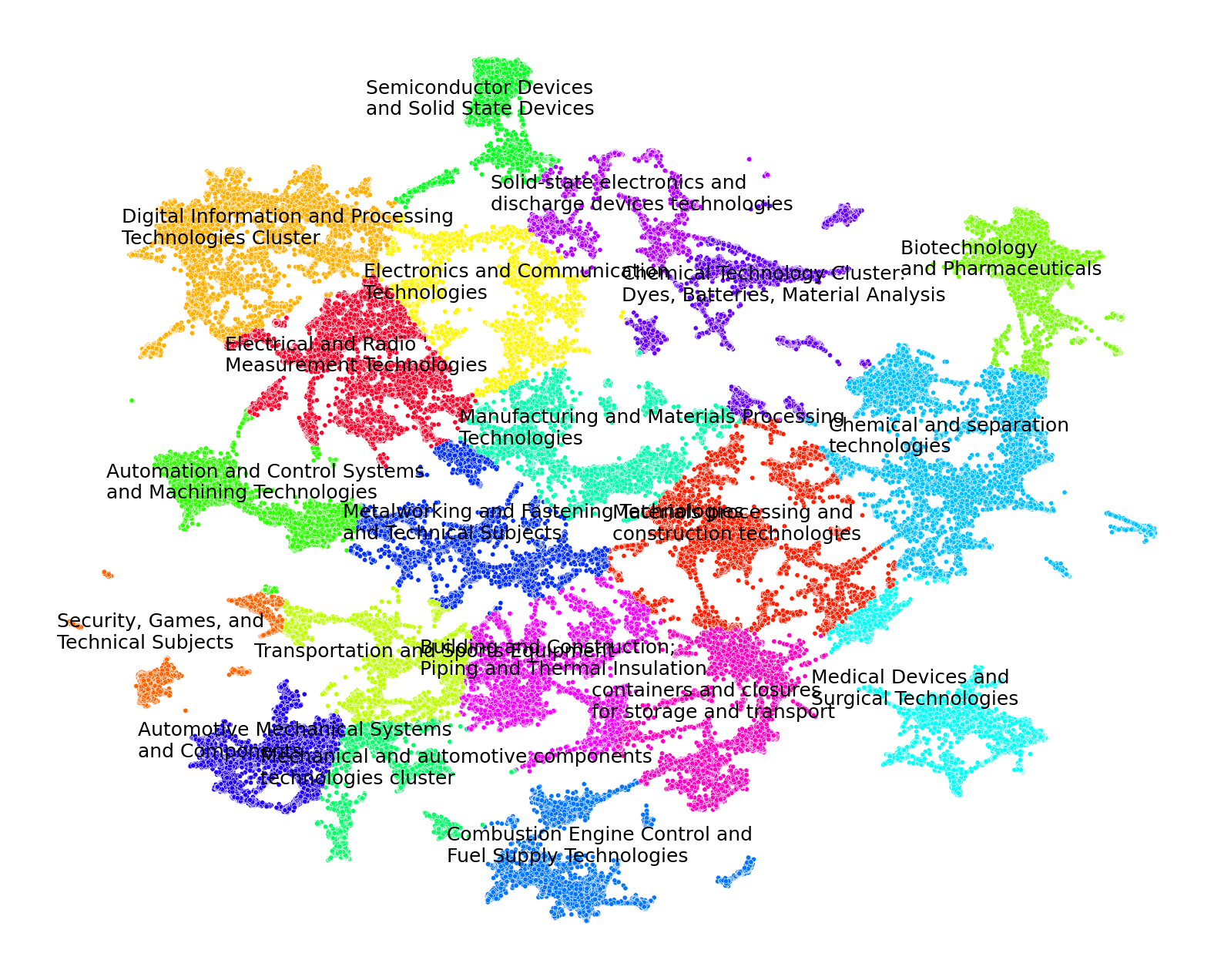}
\caption{Visual representation of the technology space of CPC codes. An interactive version  is available at https://complexly.github.io/gbb/}
\label{fig:techspace}
\end{figure}

For visualization purposes, we reduce the 50-dimensional CPC2vec embedding space to two dimensions using the UMAP algorithm. This yields a technology space, shown in Fig. \ref{fig:techspace}, that maps the similarities among all CPC codes. We use this visualization to depict the patent portfolios of firms and regions at the CPC level, as well as the composition of the GBBs and source fields that we identify in the coming sections.

\subsection{Identifying GBBs and their source fields}\label{sec:SI_identify_gbb}

To identify coherent sets of technology that represent frequent additions in CCMT inventions but not in their nongreen counterparts, we cluster such additions in the CPC2vec embedding space. To be precise, we implement the density-based HDBSCAN algorithm, which identifies clusters while also accommodating outliers. Following recommendations in the HDBSCAN documentation, we first reduce the dimensionality of CPC2vec vectors  for the  ``extra'' codes from the original 50 dimensions to 5 dimensions using UMAP. This dimension reduction decreases computational demands and mitigates noise in the dataset by correcting for semantic similarities among CPC codes. Next, we set the minimum cluster size parameter of the HDBSCAN algorithm to 100 to extract larger, more meaningful technological groupings. This process yields 82 GBB technology clusters. We repeat the same procedure, but now for the left ``base'' codes that are common across the CCMT patent and its nongreen matches. This yields 193 source field clusters.

In several of the analyses below, we weight CPC codes by how reliably they are identified as extra codes, Let $s^\mathrm{text}_{pb}$ denote the similarity between CCMT patent $p$ and matched nongreen patent $b$, i.e., the cosine similarity of their text embeddings (see section~\ref{sec:SI_text_similarity}). For a given CCMT patent, code $c$ can be identified as an extra code against one or several of its matched patents. Let $B_{cp}$ denote this set of matched patents. To determine a weight at the level of patent and code, we take the maximum text similarity across these matches:
\[
q_{cp}=\max_{b\in B_{cp}} s^\mathrm{text}_{pb}.
\]
To arrive at a weight at the level of the code, we average these values across the set $P_c$ of all CCMT patents on which $c$ is an extra code:
\[
q_c=\frac{1}{|P_c|}\sum_{p\in P_c} q_{cp}.
\]

\subsection{Naming GBBs and source fields}

To name each GBB, we rely on a large language model. \textcolor{red}{We use up to 30 CPC descriptions per cluster, ranked by the weight $q_c$ defined in section~\ref{sec:SI_identify_gbb}.} We feed these titles to the GPT-4o model to produce a short label  and a longer description for each GBB and source field. To do so, we use the following prompt template:

\begin{figure}[htp]
\centering
\begin{mdframed}[linecolor=black, linewidth=1pt]
Please name the cluster of technologies from the descriptions, the data is in following format:

[\{"tech":"description of technology", "weight":"weight of technology"\},
\{"tech":"description of technology2", "weight":"weight of technology2"\}]

Weight ranges from -1 to 1, higher weight value means you should put more emphasis on the corresponding description of technology.

You should only answer in the following json format:

\{"name":"name you assigned to the cluster, less than 5 words","desc":"summarized description of the cluster, less than 20 words"\}

Data:
\{clusteringresult\}

Your answer in plain text without code block syntax around it:
\end{mdframed}
\caption{Prompt template for naming clusters}
\label{fig:prompt}
\end{figure}

The results are returned in json format, yielding two properties for each GBB or source field: a short label of up to 4 words and a longer descriptions of up to 19 words.

\subsection{Structure of source-GBBs-target}\label{sec:SI_network_structure}
Tables~\ref{tab:gbb_degree_extremes} and~\ref{tab:source_degree_extremes} list the five highest- and lowest-degree GBBs and source fields in the original network.

\begin{table}[htbp]
\centering
\small
\setlength{\tabcolsep}{4pt}
\caption{The five highest- and lowest-degree GBBs. Degree counts source--GBB connections with $RCA>1$ in the original network of 82 GBBs and 193 source fields. Cluster IDs identify the original groups; equal degrees are ties.}
\label{tab:gbb_degree_extremes}
\begin{tabular}{@{}p{0.39\linewidth}r p{0.39\linewidth}r@{}}
\toprule
Highest degree & Degree & Lowest degree & Degree \\
\midrule
c26: Modern Construction Materials & 61 & c42: Advanced Molecular Sieves & 10 \\
c23: Advanced Measurement and Analysis & 59 & c45: Hydrocarbon Catalysis & 16 \\
c25: Advanced Medical Devices & 58 & c4: Rotary-Piston Machinery & 17 \\
c33: Advanced Laminate Technologies & 55 & c41: Advanced Catalyst Materials & 18 \\
c12: Advanced Composite Manufacturing & 53 & c80: Power Conversion Systems & 18 \\
\bottomrule
\end{tabular}
\end{table}

\begin{table}[htbp]
\centering
\small
\setlength{\tabcolsep}{4pt}
\caption{The five highest- and lowest-degree source fields. Degree counts source--GBB connections with $RCA>1$ in the original network of 82 GBBs and 193 source fields. Cluster IDs identify the original groups; equal degrees are ties.}
\label{tab:source_degree_extremes}
\begin{tabular}{@{}p{0.39\linewidth}r p{0.39\linewidth}r@{}}
\toprule
Highest degree & Degree & Lowest degree & Degree \\
\midrule
s114: Piezoelectric and Electromechanical Tech & 29 & s58: Advanced Valve Gear Systems & 4 \\
s121: Cooking and Heating Systems & 27 & s174: Electronic Clutch Control & 4 \\
s113: Advanced Chromatography Techniques & 26 & s151: Advanced Wheel Technologies & 4 \\
s108: Advanced Lasers Technology & 25 & s6: Advanced Battery Technologies & 5 \\
s35: Drying Technologies & 25 & s173: Advanced Vehicle Control Systems & 5 \\
\bottomrule
\end{tabular}
\end{table}

\begin{figure}[htp]
     \centering
     \begin{subfigure}[b]{0.45\textwidth}
         \centering
         \includegraphics[width=\textwidth]{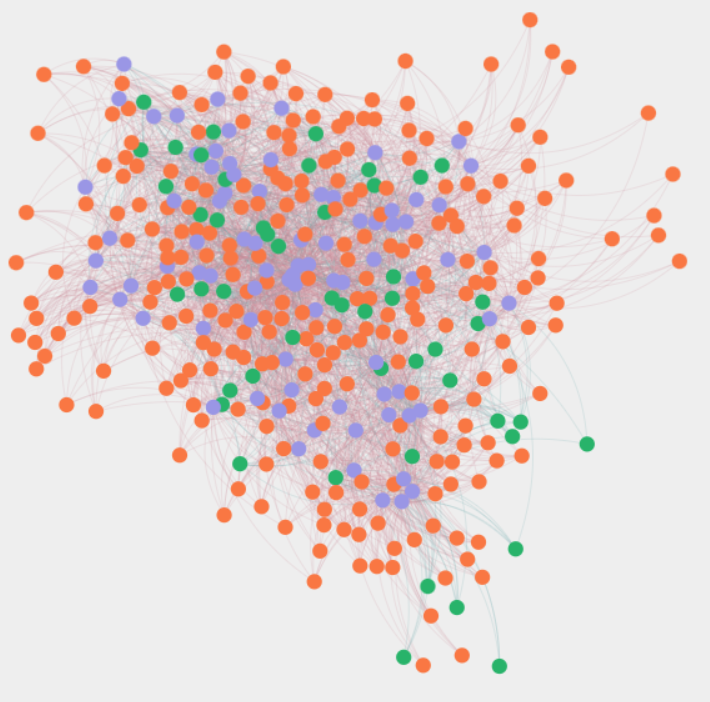}
         \caption{Force directed layout of source-GBB-target}
         \label{fig:netviz}
     \end{subfigure}
     \hfill
     \begin{subfigure}[b]{0.45\textwidth}
         \centering
         \includegraphics[width=\textwidth]{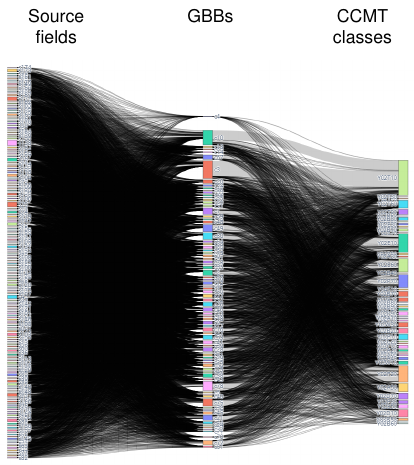}
         \caption{Sankey diagram of source-GBB-target}
         \label{fig:sankey}
     \end{subfigure}
        \caption{Visualization of relationship between source-GBB-target. Panel~(\subref{fig:sankey}) gives an overview; individual technologies and links can be explored in the interactive visualization at \url{https://complexly.github.io/gbb/}.}
        \label{fig:src-gbb-trg}
\end{figure}

Figure~\ref{fig:src-gbb-trg} plots the tripartite network that connects sources to GBBs and GBBs to CCMT technologies, once using a force-directed layout and once as a Sankey diagram. These networks, as well as the  GBBs and source fields can be explored in detail in an interactive visualization at \url{https://complexly.github.io/gbb/}. 

\begin{revision}
\subsection{Robustness of GBB anatomy under alternative patent matching and clustering}\label{sec:SI_gbb_stability}

\subsubsection{Matching patents with embedding models trained on patent-specific tasks}

In section~\ref{sec:SI_vali_human}, we tested how well match candidates are ranked by our text similarity scores derived from FSE embeddings. Here, we replace the text-embedding model itself with two embedding models trained on patent-specific tasks: PaECTER and PatEmbed. PaECTER is a version of BERT for Patents that was fine-tuned on examiner-added citations to align embeddings of patents linked by a citation \cite{ghosh_paecter_2024}. PatEmbed-base (henceforth, ``PatEmbed'') belongs to the patembed family of patent-text embedding models, which are trained jointly on 13 patent-specific retrieval, classification and paraphrase tasks \cite{ayaou_patenteb_2025}. We use the archived embeddings of PatEmbed generated without task-specific prompts. We find that the choice of embedding model matters a great deal at the start of our pipeline, but progressively less at each subsequent step.

Table~\ref{tab:functional_embedding_validation} shows how often either embedding yields the same closest match as the FSE embeddings used in the main text, as well as how many extra codes are on average identified in the CCMT patent. PaECTER selects the same top nongreen comparison patent in 7.0\%, PatEmbed  in 6.3\% of cases. Although these percentages are orders of magnitude larger than expected by chance --- given that there are millions of potential matches --- the two embeddings very often pick a different match for a CCMT patent. However, in terms of the average number of added CPC codes that are subsequently identified, the alternative models are much closer to the FSE-based results (third column of Table~\ref{tab:functional_embedding_validation}).

\begin{table}[htb]
\centering
\caption{Comparison of text-embedding models. \emph{Same top match as FSE} is calculated over all 288,579 CCMT patent families. \emph{Mean extra codes per family} is the average number of distinct extra CPC codes per CCMT patent family.}
\label{tab:functional_embedding_validation}
\small
\begin{tabular}{lcc}
\toprule
Model & \shortstack{Same top match\\as FSE} & \shortstack{Mean extra codes\\per family}\\
\midrule
FSE & 100\% & 0.534 \\
PaECTER & 7.0\% & 0.596 \\
PatEmbed & 6.3\% & 0.594 \\
\bottomrule
\end{tabular}
\end{table}

\subsubsection{From different patent matches to recurring extra technologies}\label{sec:SI_extra_code_overlap}

How does the fact that the alternative embedding models tend to select different matched patents affect which CPC codes are identified as extra? To answer this question, we need to process the pairings from each embedding in the next steps of our pipeline.

For CCMT patent $i$, let $S_i^F$ denote the set of full CPC codes identified as extra in at least one valid match of $i$ with FSE and $S_i^A$ the corresponding set obtained with an alternative embedding. We measure their overlap with the Jaccard score
\[
J_i=\frac{|S_i^F\cap S_i^A|}{|S_i^F\cup S_i^A|}.
\]
A value of 1 means that the sets are identical, while 0 means that they have no codes in common.
We calculate this once for the extra codes and once for all CPC codes on the matched nongreen patents. In the former case, we use detailed CPC codes and all CCMT patents with at least one extra code under either model. In the latter case, we aggregate codes to four-character CPC level (i.e. subclass level) and use all CCMT patents with a valid match under either model. A CCMT patent with codes under only one model receives a Jaccard score of zero. The average across CCMT patents is 0.347 (PaECTER) and 0.348 (PatEmbed) for codes on matched nongreen patents, and 0.338 and 0.333 for extra codes.

Next, we look at how often a CPC code is added as an extra code across the hundreds of thousands of comparisons. Let $n_c^F$ and $n_c^A$ be the counts of CCMT patents to which extra code $c$ is assigned under FSE and under an alternative model, $A$. We compare these frequency profiles using their cosine similarity:
\[
C(n^F,n^A)=\frac{\sum_c n_c^F n_c^A}
{\sqrt{\sum_c(n_c^F)^2}\sqrt{\sum_c(n_c^A)^2}}.
\]
Using either alternative embedding yields a cosine similarity very close to 1 (PaECTER: 0.981,  PatEmbed: 0.977). Thus, although the sets of comparison patents are very different across embedding models, the resulting aggregate relative frequencies of extra codes remain all but the same.

\subsubsection{From recurring technologies to GBBs and the source--GBB network}\label{sec:sensitivity_embeddings_endofpipeline}

We next compare how the recurring codes are grouped into GBBs. Compared to the existing (FSE-based) run, GBBs can split into several GBBs or multiple GBBs may merge into one and GBBs may change their content. In fact, with the alternative embedding models, we obtain 95 (PaECTER) and 98 (PatEmbed) GBBs, compared to 82 with FSE. We therefore focus on the content (CPC-code composition) of GBBs. Because extra codes can come from any of the up to 10 potential matches, and because matches differ in their quality, we weight each code by a run-$r$-specific version of $q_c$ (see section~\ref{sec:SI_identify_gbb}), $q_c^r$.

We now use these weights to calculate a weighted correspondence between an FSE-based GBB, $f$, and an alternative GBB, $a$, where $f$ and $a$ represent the sets of CPC codes associated with a given GBB in their run:
\[
W_{fa}=\sum_{c\in f\cap a}q_c^Fq_c^A.
\]
A shared code therefore receives more weight when it came from matched patents with high similarity scores in both runs. We normalize the correspondence weights to $p_{fa}=W_{fa}/\sum_{f,a}W_{fa}$ and calculate normalized mutual information (NMI):
\[
\operatorname{NMI}(F,A)=
\frac{\sum_{f,a}p_{fa}\log[p_{fa}/(p_f p_a)]}
{\sqrt{H(p_f)H(p_a)}} ,
\qquad H(p)=-\sum p\log p,
\]
where $p_f=\sum_a p_{fa}$ and $p_a=\sum_f p_{fa}$ and cells with $p_{fa}=0$ contribute zero. We run this analysis only for CPC codes that end up in GBB clusters in both runs. Extra CPC codes that drop out during the pipeline's HDBSCAN clustering step in either run are omitted. NMI ranges from 0 to 1. A value of 1 means that knowing a code's FSE-based GBB perfectly identifies its alternative GBB, and vice versa. 

Resulting NMIs are very high: 0.951 for PaECTER and 0.953 for PatEmbed. To compare this to a random benchmark, we take the mapping of extra CPC codes to GBB clusters in the FSE run and one of the alternative runs, using only CPC codes that occur in both runs. Next, we permute the alternative run's cluster labels across CPC codes 1,000 times, preserving the number of codes in each alternative cluster and keeping each code's $q_c^Fq_c^A$ weight fixed. The comparisons contain 22,507 shared codes for PaECTER and 23,013 for PatEmbed. The mean NMI across 1,000 shuffles is 0.04270 for PaECTER (central 95\% permutation interval 0.04150--0.04393) and 0.04267 for PatEmbed (0.04153--0.04390). The observed NMIs lie far outside these intervals.

To assess the role that mergers and splits of GBBs play, we also calculate for each FSE GBB the share of correspondence weight captured by the two most closely connected alternative GBBs, and vice versa, taking the average  across the two directions. Let $B_2(f)$ contain the two alternative blocks with the largest $W_{fa}$ for FSE block $f$, and let $A_2(a)$ contain the two strongest FSE counterparts of alternative block $a$. Then
\[
T_2=\frac{\sum_f\sum_{a\in B_2(f)}W_{fa}
+\sum_a\sum_{f\in A_2(a)}W_{fa}}
{2\sum_{f,a}W_{fa}}.
\]
If a GBB has only one counterpart, we use that counterpart. Averaged across the two directions, the two strongest counterparts account for 97.2\% of the weighted overlap in CPC codes for PaECTER and 97.0\% for PatEmbed. This implies that the change from 82 FSE-based GBBs to 95 and 98 alternative GBBs consists mainly of splits and mergers of the same broad technological blocks.

Finally, we compare the source--GBB networks across runs. Let $L_{sg}$ count distinct CCMT patent families containing at least one code assigned to source field $s$ and at least one assigned to GBB $g$. We write $L^F_{sf}$ for these counts in the FSE-based run and $L^A_{sa}$ in an alternative run. These counts underlie the source--GBB network of the main text, in which a source field and a GBB are linked if their RCA exceeds 1. Throughout this comparison, we keep the 193 source fields of the main analysis fixed, leaving varying source fields for section~\ref{sec:SI_source_stability}.

Because the alternative models yield different GBBs, we first express their patent counts in terms of the original 82 GBBs. We divide the patents that connect source field $s$ to an alternative GBB $a$ among the FSE-based GBBs $f$ in proportion to how strongly $a$ corresponds to each of them:
\[
\tilde{L}_{sf}=\sum_a L^A_{sa}\,\frac{W_{fa}}{\sum_{f'}W_{f'a}}.
\]
If an alternative GBB corresponds to a single FSE-based GBB, all its patents are thus attributed to that GBB. Patents of alternative GBBs without any FSE-based counterpart are collected in a separate category, for which the FSE-based counts are zero. We then compare $\tilde{L}_{sf}$ to $L^F_{sf}$ across all pairs of source fields and GBBs, using the cosine similarity introduced in section~\ref{sec:SI_extra_code_overlap}. 

The similarities are again very high: 0.960 for PaECTER and 0.956 for PatEmbed. If we instead compare RCA values, rescaled as $RCA/(1+RCA)$ such that they lie between 0 and 1, the similarities are 0.936 and 0.935. Moreover, the different embedding models also yield highly similar ubiquity rankings for GBBs. We define a GBB's ubiquity as the number of source fields in which it has an RCA above 1, i.e., its degree in the source--GBB network. For the alternative models, we calculate RCA values from $\tilde{L}_{sf}$. The rank correlations between the resulting ubiquities and those in the FSE-based network are 0.823 for PaECTER and 0.863 for PatEmbed.

\begin{table}[htb]
\centering
\caption{Stability across the GBB pipeline using alternative patent text embeddings. The first four rows show that the comparison patents and their extra codes change substantially but that the aggregate extra-code frequency distribution does not. The remaining rows show that the GBBs themselves and the source--GBB network also remain closely aligned.}
\label{tab:retrieval_gbb_stability}
\footnotesize
\begin{tabular}{lcc}
\toprule
Comparison with FSE-based result & PaECTER & PatEmbed \\
\midrule
Same closest nongreen match  & 7.0\% & 6.3\% \\
Mean Jaccard score (codes on matched nongreen patents) & 0.347 & 0.348 \\
Mean Jaccard score (extra codes) & 0.338 & 0.333 \\
Extra-code frequency distribution, cosine similarity & 0.981 & 0.977 \\
\addlinespace
GBB count & 95 & 98 \\
GBB NMI & 0.951 & 0.953 \\
Top-two correspondence share & 0.972 & 0.970 \\
Source--GBB patent count distribution, cosine similarity & 0.960 & 0.956 \\
Source--GBB RCA distribution, cosine similarity & 0.936 & 0.935 \\
GBB ubiquity-rank Spearman correlation & 0.823 & 0.863 \\
\bottomrule
\end{tabular}
\end{table}

Table~\ref{tab:retrieval_gbb_stability} summarizes these comparisons. The choice of embedding model has a large impact at the start of our pipeline: the alternative models almost always select a different closest match. Moreover, for individual CCMT patents, they often identify different extra codes. However, these differences fade as we move along the pipeline to the more aggregate levels of GBBs and the source--GBB network. The frequency distribution of extra codes is similar and these codes  are grouped into very similar GBBs. These GBBs, in turn, connect to source fields in very similar ways. This is what the pipeline was designed to do: reduce noise in matches and CPC codes. As a result, GBBs and the source--GBB network emerge as robust features that do not depend on the specific embedding model used to match patents.

\begin{figure}[htbp]
\centering
\includegraphics[width=\textwidth]{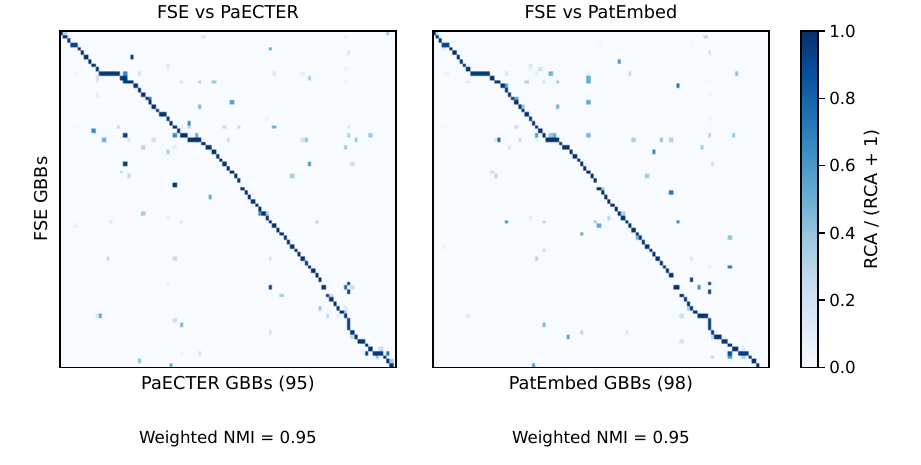}
\caption{Correspondence between GBBs obtained with FSE and with PaECTER (left) or PatEmbed (right). Rows are the 82 FSE-based GBBs; columns the GBBs obtained with the alternative embedding model, ordered by the FSE-based GBB to which they correspond most strongly. Colors show how strongly each pair of GBBs corresponds, based on the weights $W_{fa}$, expressed as an RCA value and rescaled as $RCA/(1+RCA)$. The concentration of high values along the diagonal shows that the alternative embedding models recover largely the same GBBs. High values off the diagonal indicate GBBs that are split or merged.}
\label{fig:retrieval_gbb_correspondence}
\end{figure}

\subsubsection{Sensitivity to modeling choices in GBB identification}\label{sec:sensitivity_GBBs}

The next step in the pipeline is the construction of GBBs. GBBs are essentially groupings of CPC codes, which are represented as the 50-dimensional vectors from our CPC2vec procedure. In the main text, these groupings are generated using HDBSCAN. Because clustering algorithms (including HDBSCAN) do poorly in high dimensions, we first reduce the dimensionality of the CPC space to 5 dimensions, using UMAP (15 neighbours). Next, we use HDBSCAN (minimum cluster size of 100). The extra codes that enter this step are identified as described in the Methods section: we use up to 10 nongreen matches per CCMT patent and select codes whose distance score lies more than 1.5 interquartile ranges above the third quartile. Below, we vary each of these choices in turn. In these sensitivity analyses, we will keep the source partitions fixed and only focus on how GBBs are affected.

Table~\ref{tab:parameter_stability} reports results. In each row, we varied the clustering setting named in the first column and held the other four settings at their original values. The remaining columns compare the resulting GBBs to the ones used in the main text, using four of the measures of Table~\ref{tab:retrieval_gbb_stability}: the number of GBBs (``\# GBBs''), the GBB NMI (``GBB content''), the cosine similarity of source--GBB patent counts (``network'') and the rank correlation of GBB ubiquity (``GBB ubiq.''). As in section~\ref{sec:sensitivity_embeddings_endofpipeline}, we first express the source--GBB patent counts of each alternative run in terms of the original GBBs.

\begin{table}[htb]
\centering
\caption{Stability of GBBs and the source--GBB network when each modeling choice is changed separately. Each row changes only the stated setting and holds all others at their original values. In each row, results are compared to the original ones.}
\label{tab:parameter_stability}
\footnotesize
\begin{tabular}{p{0.43\textwidth}rrrr}
\toprule
Single setting changed & \# GBBs & GBB content & network & GBB ubiq. \\
\midrule
UMAP neighbours: 5 & 84 & 0.955 & 0.940 & 0.860 \\
UMAP neighbours: 10 & 80 & 0.977 & 0.972 & 0.891 \\
UMAP neighbours: 30 & 79 & 0.977 & 0.982 & 0.874 \\
UMAP neighbours: 50 & 77 & 0.984 & 0.981 & 0.855 \\
\addlinespace
UMAP dimensions: 2 & 91 & 0.966 & 0.950 & 0.905 \\
UMAP dimensions: 10 & 81 & 0.979 & 0.976 & 0.879 \\
\addlinespace
HDBSCAN minimum cluster size: 50 & 168 & 0.933 & 0.908 & 0.913 \\
HDBSCAN minimum cluster size: 75 & 112 & 0.961 & 0.967 & 0.941 \\
HDBSCAN minimum cluster size: 150 & 53 & 0.955 & 0.940 & 0.805 \\
HDBSCAN minimum cluster size: 200 & 41 & 0.944 & 0.922 & 0.592 \\
\addlinespace
Extra-code IQR multiplier: 1.00 & 107 & 0.965 & 0.953 & 0.887 \\
Extra-code IQR multiplier: 1.25 & 88 & 0.973 & 0.972 & 0.871 \\
Extra-code IQR multiplier: 1.75 & 69 & 0.964 & 0.966 & 0.793 \\
Extra-code IQR multiplier: 2.00 & 53 & 0.948 & 0.951 & 0.696 \\
\addlinespace
Nongreen matches: 5 & 61 & 0.967 & 0.963 & 0.843 \\
Nongreen matches: 8 & 74 & 0.983 & 0.986 & 0.775 \\
Nongreen matches: 15 & 98 & 0.963 & 0.963 & 0.891 \\
Nongreen matches: 20 & 97 & 0.971 & 0.977 & 0.933 \\
\bottomrule
\end{tabular}
\end{table}

\begin{figure}[p]
\centering
\includegraphics[width=0.98\textwidth]{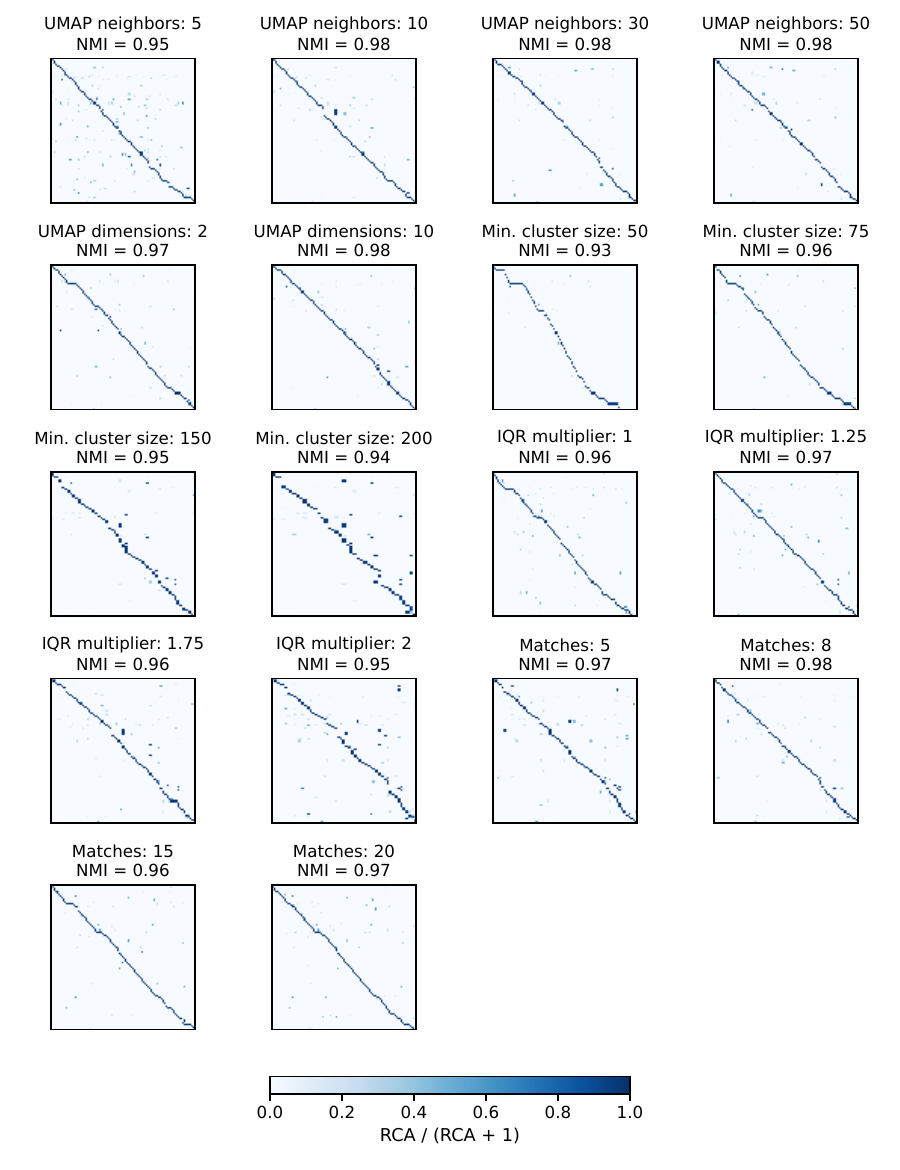}
\caption{Correspondence between the original GBBs and the GBBs obtained under alternative modeling choices. Each panel compares the 82 original GBBs (rows) with the GBBs of one alternative run (columns), ordered by the original GBB to which they correspond most strongly. Each of the 18 panels changes one setting, following the order in Table~\ref{tab:parameter_stability}. Colors show the correspondence between GBBs, expressed as RCA/(1 + RCA). The concentration of high values along the diagonal shows that largely the same GBBs are recovered when modeling choices change.}
\label{fig:clustering_parameter_alignment}
\end{figure}

The exact UMAP settings matter little: the number of GBBs stays between 77 and 91, the GBB NMI is at least 0.955 and the cosine similarity of source--GBB patent counts at least 0.940.

The minimum cluster size in HDBSCAN has, as expected, a large effect on the number of GBBs, which ranges from 168 (minimum cluster size of 50) to 41 (minimum cluster size of 200). Nevertheless, the GBB NMI remains at least 0.933 and the cosine similarity of source--GBB patent counts at least 0.908. That is, the original GBBs are mostly split up or merged, not reshuffled.

Changing the threshold for extra codes or the number of nongreen matches also affects the number of GBBs, but the GBB NMI remains high (0.948--0.983), as does the cosine similarity of source--GBB patent counts (0.951--0.986).

The ubiquity ranking of GBBs is more sensitive, with rank correlations between 0.592 and 0.941. The lowest value occurs when the minimum cluster size is doubled to 200, which reduces the number of GBBs from 82 to 41. This is unsurprising: when small, poorly connected GBBs are merged with better connected ones, their ubiquity ranks change. In line with this, all rank correlations below 0.85 occur in runs that yield fewer than 75 GBBs. The lower correlations are therefore consistent with sensitivity to coarser aggregation.

\subsubsection{Robustness to source-field clustering}\label{sec:SI_source_stability}

So far, we have kept the 193 source fields of the main text fixed. Here, we instead keep the original 82 GBBs fixed and vary how we identify source fields. Like GBBs, source fields are clusters of CPC codes obtained through a combination of UMAP and HDBSCAN (see Methods). We run three alternative pipelines. The first lowers HDBSCAN's minimum cluster size from 100 to 50. This yields more and smaller source fields. The second raises this setting to 200, which yields fewer and larger ones. The third increases the number of UMAP neighbours from 15 to 30. The remainder of the pipeline stays at its original settings.\footnote{By default, HDBSCAN's \texttt{min\_samples} is set equal to the minimum cluster size. We override this such that this parameter remains fixed at 100 as in the original analysis.}

To compare source--GBB patent counts, we follow the procedure of section~\ref{sec:sensitivity_embeddings_endofpipeline}, with the roles of GBBs and source fields reversed. There, we had to redistribute the patent counts of alternative GBBs over the original GBBs. To do so, we used weights, $W_{fa}$, that express how strongly original GBB $f$ corresponds to alternative GBB $a$. Here, we need to redistribute patent counts of alternative sources over the original sources. To do so, we calculate the same kind of weights, $W_{su}$, to express how strongly original source field $s$ corresponds to alternative source field $u$, and redistribute the patent counts of each alternative source field across the original source fields:
\[
\tilde{L}_{sf}=\sum_{u} L^{A}_{uf}\,\frac{W_{su}}{\sum_{s'}W_{s'u}},
\]
where $L^{A}_{uf}$ is the number of CCMT patents that connect alternative source field $u$ to GBB $f$. As before, we then compare $\tilde{L}_{sf}$ to the original counts $L_{sf}$ using the cosine similarity of patent counts, and report NMIs between original and alternative source fields. The ubiquity of GBBs in terms of how they connect to alternative source fields can be compared directly, because the GBBs do not change across runs.

Table~\ref{tab:source_stability} reports results. Lowering the minimum cluster size or increasing the number of UMAP neighbours changes little: the NMI between original and alternative source fields  is 0.992 in both cases (column ``source content''), the cosine similarity of source--GBB patent counts is 0.988 and 0.968 (column ``network''), and the ubiquity ranking of GBBs is nearly unchanged (rank correlations of 0.989 and 0.984, column ``GBB ubiq.''). Raising the minimum cluster size to 200 has a larger effect, reducing the number of source fields from 193 to 107 (column ``\# sources''). Even so, the NMI remains 0.956, the cosine similarity of patent counts 0.871 and the rank correlation of GBB ubiquity 0.851. Which GBBs are widely applicable and which are specialized therefore depends little on how exactly source fields are generated.

\begin{table}[htbp]
\centering\footnotesize
\setlength{\tabcolsep}{4pt}
\caption{Stability of the source--GBB network with alternative source fields. Each row changes only the stated setting and holds all others, including the 82 GBBs, at their original values. In each row, results are compared to the original ones.}\label{tab:source_stability}
\begin{tabular}{p{0.32\linewidth}rrrr}
\toprule
Setting changed & \# sources & source content & network & GBB ubiq. \\
\midrule
Minimum cluster size: 50 & 209 & 0.992 & 0.988 & 0.989 \\
Minimum cluster size: 200 & 107 & 0.956 & 0.871 & 0.851 \\
UMAP neighbours: 30 & 190 & 0.992 & 0.968 & 0.984 \\
\bottomrule
\end{tabular}
\end{table}

\subsection{Sensitivity to CPC aggregation level}\label{sec:SI_cpc_granularity}
So far, the CPC codes we used are \emph{subgroup} classes, the most detailed level available. The disadvantage of using such highly disaggregated codes is that distinctions among them are sometimes hard to appreciate, such that several codes would be equally suitable to label a given patent. This can lead to misclassification.

Here, we therefore aggregate these subgroups to \emph{main groups}. For instance, the subgroup B60L50/60 (electric propulsion with power supplied by batteries) belongs to the main group B60L50/00 (electric propulsion with power supplied within the vehicle). We will refer to subgroups as detailed codes and to main groups as coarse codes.

Note, however, that aggregation may also come at a cost. First, relatedness between the classes of a classification system typically does not closely follow the system's hierarchy \cite{bryce2009general,neffke2013skill,hidalgo2007product}. Using coarser codes may therefore obscure real heterogeneity and lead to spurious connections across classes. Second, some added technological components may be lost: detailed additional codes on a CCMT patent that do not occur on the matched nongreen patents may share coarser hierarchical parent classes with codes on these matched patents. In this case, these additions will be overlooked.

To test how aggregation affects our results, we repeat the analysis with coarse codes. We keep the patent matches of the main analysis and replace detailed CPC codes  by their corresponding coarse codes. We then follow the same procedure as before: we train a CPC2vec model (now on coarse codes), identify extra codes, and group them into GBBs using UMAP and HDBSCAN. Settings are kept the same as in the original analysis, with the exception of HDBSCAN's minimum cluster size. Because there are far fewer coarse extra codes to cluster than detailed ones (4,454 against 37,801), we lower minimum cluster size proportionally, from 100 to 12.

As expected, aggregation leads to a loss of extra codes. If we aggregate the extra codes of the main analysis to coarse codes, we obtain 116,245 combinations of a CCMT patent and a coarse extra code. Basing the pipeline on coarse codes, our procedure instead identifies 64,361 such combinations. However, this reduction does not translate into a change in the substantive nature of the extra codes we identify: the two frequency distributions of coarse codes ---once aggregated from our original analysis and once generated from scratch by running the pipeline at a more aggregate level---  is similar (cosine similarity of 0.869).

The analysis with aggregated CPC codes yields 105 GBBs, against the original 82 GBBs. To compare the two sets of GBBs, we link them through their codes: an original GBB and a GBB based on coarse codes are linked whenever a detailed code in the former falls under a coarse code in the latter. The more such codes two GBBs have in common, the closer their correspondence. As in section~\ref{sec:sensitivity_embeddings_endofpipeline}, we weight codes by how often and how reliably they are identified as extra codes. 

The results are shown in Table~\ref{tab:cpc_main_group_stability}. The NMI between the two sets of GBBs is 0.837, and 91.9\% of the correspondence between them is accounted for by each GBB's two closest counterparts. Figure~\ref{fig:cpc_main_group_alignment} furthermore shows the correspondence for all GBBs. Although, supporting the notion that aggregation bears some costs,  the correspondence is weaker than in the preceding sections, we recover structurally similar GBBs.

\begin{table}[htbp]
\centering\footnotesize
\caption{Stability of extra codes and GBBs after coarsening CPC codes. Rows are defined as in Table~\ref{tab:retrieval_gbb_stability}. }
\label{tab:cpc_main_group_stability}
\begin{tabular}{lr}
\toprule
Comparison with original result & Coarse CPC codes \\
\midrule
Extra-code frequency distribution, cosine similarity & 0.869 \\
GBB count & 105 \\
GBB NMI & 0.837 \\
Top-two correspondence share & 0.919 \\
\bottomrule
\end{tabular}
\end{table}

\begin{figure}[p]
\centering
\includegraphics[width=0.4\textwidth]{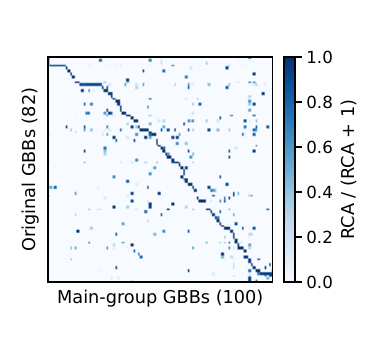}
\caption{Correspondence between the original GBBs and GBBs based on coarse CPC codes. Rows are the 82 original GBBs, columns the 100 of 105 GBBs based on coarse codes that can be linked to an original GBB. GBBs based on coarse codes are ordered by the original GBB to which they correspond most strongly. Colors show strength of correspondence, measured by RCA/(1 + RCA).}
\label{fig:cpc_main_group_alignment}
\end{figure}

\end{revision}

\section{Alternative methods to extract GBBs}\label{sec:SI_alternative_methods}

In this section, we explore two alternative ways to reduce the impact of noisy CPC code assignments in our identification of  GBBs. The first changes how we cluster extra codes, replacing the HDBSCAN step of section~\ref{sec:SI_GBB_construction} \begin{revision} by $k$-means clustering.\end{revision} The second takes a radically different approach to the problem, starting from how we define ``extra'' technology codes.

\subsection{k-means}

\begin{revision}The k-means comparison uses the same 5-dimensional UMAP coordinates as HDBSCAN. This isolates the impact of the clustering algorithm. We fit mini-batch $k$-means, a variation designed to handle large datasets efficiently, with 82 requested centers, random state 0, batch size 10 and automatic initialization. We do so weighting each CPC code by how often it occurs as an extra code in our data. This yields 78 non-empty GBB clusters. Next, we assess how well these new GBBs map onto the original GBBs, following the approach of section~\ref{sec:sensitivity_embeddings_endofpipeline}.\end{revision}

Figure~\ref{fig:comparekm2} compares these GBBs to the original ones, showing the original GBBs from HDBSCAN  on the horizontal axis and the new GBBs from k-means along the vertical axis. We sort both axes to put GBBs on one axis in a similar position as closely connected GBBs on the other axis. To do so, we construct a bipartite network that connects HDBSCAN-based GBBs to k-means-based GBBs. Next, we use a Louvain community detection algorithm to identify communities that contain both types of GBBs and order both axes by the labels of these communities. This way, splits and mergers of GBBs appear as blocks along the diagonal.

As in Figures~\ref{fig:retrieval_gbb_correspondence} and~\ref{fig:clustering_parameter_alignment}, the concentration of entries along the diagonal shows that the two approaches yield very similar GBBs. \begin{revision}To quantify the strength of this correspondence, we weight CPC codes and calculate the NMI as in section~\ref{sec:sensitivity_embeddings_endofpipeline}. We do so across the 30,675 CPC codes that were assigned to one of the original GBBs. The NMI between the original GBBs and the GBBs derived with k-means clustering is 0.874, against 0.026 for the random benchmark described in that section.\end{revision}

\subsection{Finding orthogonal components instead of added codes}
Our second approach replaces a large part of our pipeline to identify GBBs. Rather than identifying specific, discrete, CPC codes and subsequently clustering them, here we transform both codes and patents into continuous vector representations and perform calculations within this continuous space. This yields a completely different way to handle noise. Because it changes both how we identify what CCMT patents add and how we group these additions, it is the most demanding of our robustness checks. If, in the end, we nevertheless extract a similar set of GBBs, this would suggest that GBBs reflect structure in the data and are not an artifact of how we extract them. Below, we describe the new procedure step-by-step.

\subsubsection{Converting technology codes of each patent into a continuous representation}

As a first step, we convert the discrete CPC codes of each patent into a continuous representation by summing the 50-dimensional embedding vectors of all CPC codes listed on the patent. This is illustrated in Fig.~\ref{fig:pat2vec}.

\subsubsection{Extracting orthogonal components of green patents}

\begin{figure}[htp]
    \centering
    \begin{subfigure}[b]{0.35\textwidth}
        \centering
        \includegraphics[width=\textwidth]{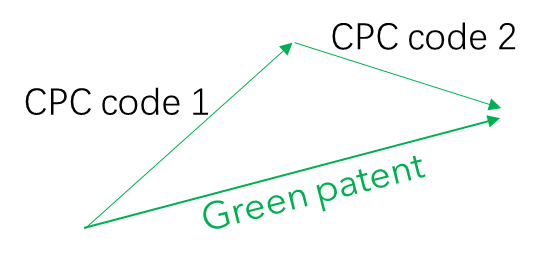}
        \caption{Convert patent to a vector by summing CPC vectors}
        \label{fig:pat2vec}
    \end{subfigure}
    \hfill
    \begin{subfigure}[b]{0.45\textwidth}
        \centering
        \includegraphics[width=\textwidth]{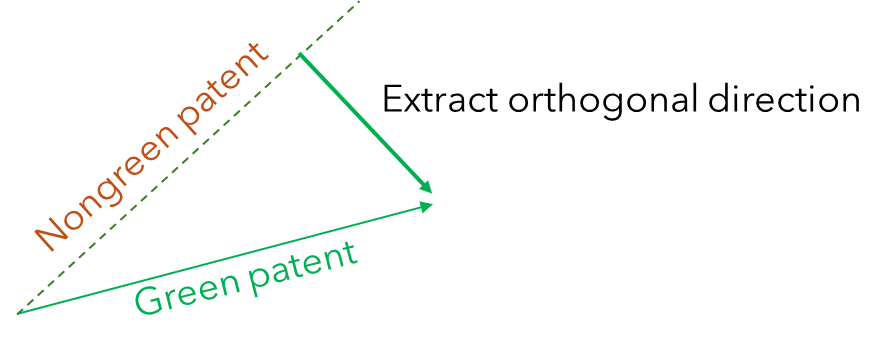}
        \caption{Extract the orthogonal component}
        \label{fig:projection}
    \end{subfigure}
       \caption{Illustration of how ``extra'' parts are extracted from CCMT patents in continuous space}
       \label{fig:extract}
\end{figure}

Next, for each CCMT--nongreen patent pair, we project the vector of the CCMT patent onto the vector of the nongreen patent. We then extract the component orthogonal to the nongreen patent  as the ``extra'' part that the CCMT patent adds to it (see Fig.~\ref{fig:projection}). Note that the extra part  now does not consist of a set of discrete CPC codes, but of a  continuous 50-dimensional vector representation of the technological addition in the CCMT patent. 

To be precise, let $\mathbf{g}$ be a 50-dimensional column vector representing the technology codes of a CCMT patent and  $\mathbf{n}$ an analogous vector representing the technology codes of a matched nongreen patent. Now, we decompose $\mathbf{g}$ into two parts: 

\begin{itemize}
    \item the projection onto $\mathbf{n}$: $\mathbf{g}_\mathrm{proj} = \frac{\mathbf{n} \mathbf{n}^T}{\mathbf{n}^T \mathbf{n}} \mathbf{g}$, and 
    \item the orthogonal component: $\mathbf{g}_\mathrm{orth} = \left( \mathbf{I} - \frac{\mathbf{n} \mathbf{n}^T}{\mathbf{n}^T \mathbf{n}} \right) \mathbf{g}$,
\end{itemize}
where $\mathbf{I}$ is the identity matrix.

\subsubsection{Clustering orthogonal components}

To find common additions on CCMT patents, we cluster the orthogonal components of all CCMT--nongreen patent pairs. Each cluster represents a generalized recurring addition, i.e., a GBB. As before, we first reduce the orthogonal components from their original 50 dimensions to 5 using UMAP and then apply HDBSCAN. Because HDBSCAN assumes that some of the vectors it clusters may be noise, it filters out those that are not clearly connected to a coherent cluster. This yields 53 clusters. Note that we could cluster the projections in the same way to find source fields. However, to keep the exposition compact, we focus on GBBs.

\subsubsection{Recover the set of CPC codes belonging to a GBB}

We now need to associate these clusters of technological additions with concrete CPC codes to define GBBs in terms of these codes. To do so, we first determine the centroid vector of each cluster. Next, we calculate the cosine similarity between each centroid and the CPC2vec vector of each individual CPC code, dropping negative similarities. The higher this similarity, the more strongly a CPC code is associated with the technological addition at hand. 

To determine which codes should be included in a GBB, we use an RCA-like measure: we compare a code's similarity to a cluster with the code's similarity to all clusters and the cluster's similarity to all codes. In particular, let $\text{cs}(v_{\tau},v_{b})$ be the cosine similarity between the embedding vector of CPC code $\tau$ and the centroid vector of GBB $b$. The relative similarity of $\tau$ to GBB $b$ is defined as:

\begin{equation}
    O_{\tau,b} = \frac{\text{cs}({v_{\tau},v_{b}})/\sum_{b'} \text{cs}({v_{\tau},v_{b'}})}{\sum_{\tau'}\text{cs}({v_{\tau'},v_{b}})/\sum_{\tau'',b''} \text{cs}({v_{\tau''},v_{b''}})}
\end{equation}

To focus on technologies that are core to a specific GBB, we define a GBB as all  CPC codes for which $O_{\tau,b}>2$.

\subsubsection{Results}

\begin{figure}[htp]
     \centering
     \begin{subfigure}[b]{0.45\textwidth}
         \centering
         \includegraphics[width=\textwidth]{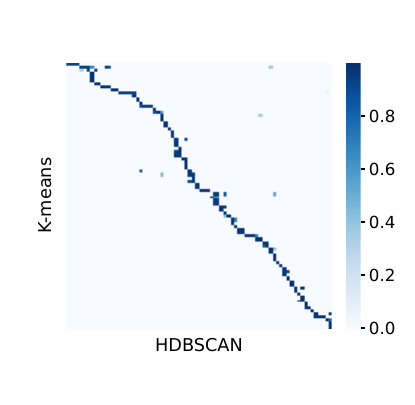}

         \caption{$k$-means clustering against HDBSCAN}
         \label{fig:comparekm2}
     \end{subfigure}
     \hfill
     \begin{subfigure}[b]{0.45\textwidth}
        \centering
        \includegraphics[width=\textwidth]{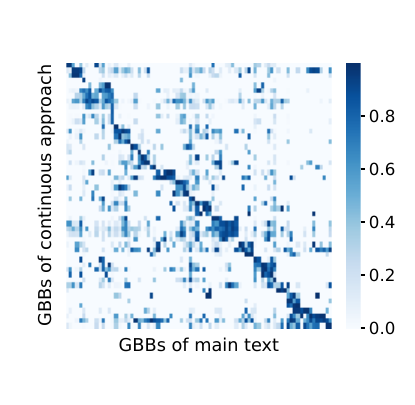}
        \caption{Discrete against continuous approach}
        \label{fig:con_dis}
     \end{subfigure}
        \caption{Comparing GBB identification approaches}
        \label{fig:comparekm}
\end{figure}

Fig.~\ref{fig:con_dis} compares the two sets. The alternative approach yields fewer and larger GBBs (53 against 82). However, these larger GBBs often consist of several closely related original GBBs: most of the entries are found around the diagonal of the matrix. In spite of the very different approach, the resulting GBBs therefore resemble the ones identified in the main text. 

\begin{revision}To quantify this resemblance, we proceed as in section~\ref{sec:sensitivity_embeddings_endofpipeline}. There, the correspondence between an original GBB $f$ and an alternative GBB $a$ was based on the CPC codes they share, where each shared code was weighted by the product of its weights in the two runs, $q_c^F q_c^A$. Here, we do the same, with one difference. In our continuous approach, CPC codes do not have a weight $q_c$. We therefore replace this weight by the cosine similarity of a code to the centroid of the corresponding new GBB. The correspondence between original GBB $f$ and new GBB $b$ thus becomes
\[
W_{fb}=\sum_{c\in f\cap b} q_c\,\text{cs}(v_c,v_b).
\]
In total, 23,248 CPC codes belong to a GBB in both approaches, 75.8\% of the 30,675 codes in the original GBBs. They cover all 82 original GBBs and all but one of the 53 new ones. Using $W_{fb}$ as before yields an NMI of 0.579. This is lower than in the previous sections. In part, this may reflect not the GBBs themselves, but a complication that arises from how we  map CPC codes to GBB centroids, because this mapping allows a single CPC code to be connected to multiple GBBs. Regardless, judged against the fact that we exchanged several components of our pipeline at once for a conceptually radically different approach, we consider the observed correspondence remarkable.\end{revision}

\begin{revision}
\section{Green entry beyond conventional technological relatedness}\label{sec:SI_entry_relatedness}

In the main text, we show that GBBs help predict which new CCMT class a firm will enter in that firms preferentially enter CCMT classes for which the firm's existing specializations cover both relevant source fields and GBBs. In fact, interaction models show that these two quantities reinforce each other's effects. 

One concern is that these findings reflect mere technological relatedness: firms tend to enter new technologies that are related to the ones they are already active in. Note that the two explanations may overlap: relevant source fields and GBBs can also be closely related to the CCMT class. The question is therefore whether source fields, GBBs and their interaction predict green entry beyond what the effect of relatedness can explain. 

To test this, we add a standard relatedness measure, relatedness density, to the entry regressions of the main text. Here, technological relatedness is inferred from how often technologies occur together on the same patents.

The entry regressions are described in the Methods section. For clarity, we provide definitions for  the two main regressors here. Let $A_{fj}=1$ if firm $f$ is specialized in GBB or source field $j$ in the base period (RCA $\geq 1$), let $L_{gt}=1$ if GBB $g$ is connected to target $t$ in the source--GBB--target network, and $L_{sg}=1$ if source field $s$ is connected to GBB $g$. Then
\begin{align*}
G_{ft}&=\sum_g L_{gt}A_{fg}, \\
S_{ft}&=\sum_s \mathbf{1}\Big\{\sum_g L_{sg}L_{gt}>0\Big\}A_{fs}.
\end{align*}
That is, $G_{ft}$ counts the relevant GBBs in which the firm is specialized, and $S_{ft}$ the source fields in which the firm is specialized that are connected to the target through at least one relevant GBB. The sample contains 600,571 firm--target observations, 12,810 firms, 49 targets and 18,823 entries.

\subsection{Constructing relatedness density}

To measure technological relatedness, we use all detailed CPC codes in sections A--H, i.e., excluding the Y codes that identify CCMT patents. This yields 206,816 codes. 

Let $N_c$ be the number of patents with CPC code $c$, $N_t$ the number with CCMT target $t$, and $N_{ct}$ the number with both. Following \cite{hidalgo2007product}, we define the relatedness between code $c$ and target $t$ as
\begin{equation}\label{eq:SI_phi}
\phi_{ct}=\min\left(\frac{N_{ct}}{N_c},\frac{N_{ct}}{N_t}\right)
=\frac{N_{ct}}{\max(N_c,N_t)}.
\end{equation}
That is, we ask how often patents with code $c$ also belong to target $t$ and vice versa, and take the smaller of the two proportions. We do so using only patents filed in the base period.

Next, let $x_{fc}$ be the number of patents of firm $f$ with code $c$ in the base period. To assess whether firm $f$ is specialized in code $c$, we once again rely on an RCA value:
\begin{equation}\label{eq:SI_entry_rca}
RCA_{fc}=\frac{x_{fc}/\sum_k x_{fk}}{\sum_h x_{hc}/\sum_{h,k}x_{hk}},
\qquad M_{fc}=\mathbf{1}\{x_{fc}>0,\ RCA_{fc}\geq1\}.
\end{equation}
Here, $h$ runs over all 14,629 firms with patents in the base period and $k$ over all codes. We then define the relatedness density of firm $f$ to target $t$ as
\begin{equation}\label{eq:SI_density}
D_{ft}=\frac{\sum_c M_{fc}\phi_{ct}}{\sum_c\phi_{ct}}.
\end{equation}
That is, density is a relatedness-weighted count of the technologies in which the firm is specialized, where the weights reflect how closely each technology is related to the CCMT class and sum to one. Density is high when the firm is specialized in many technologies closely related to the CCMT class. Firms without patents in sections A--H in the base period have no density and are dropped from the comparison.

\subsection{Entry regressions with relatedness density}

We add relatedness density to the linear probability model of the main text. With $g_{ft}=\log(1+G_{ft})$ and $s_{ft}=\log(1+S_{ft})$, we estimate
\begin{equation}\label{eq:SI_entry_controlled}
Y_{ft}=\alpha_f+\delta_t+\beta_G g_{ft}+\beta_S s_{ft}
+\beta_{GS}g_{ft}s_{ft}+\gamma_DD_{ft}+\varepsilon_{ft},
\end{equation}
where $\alpha_f$ and $\delta_t$ are firm and target (CCMT) fixed effects. Standard errors are clustered by firm and by target.

Relatedness density can be calculated for 578,417 of the 600,571 observations (96.3\%). The remaining 22,154 observations belong to 457 firms without patents in sections A--H in the base period. We estimate three models on this common sample (Table~\ref{tab:SI_entry_relatedness}): one with the density term only, one with the source-field and GBB terms only (corresponding to the result of the main text), and one with both sets of terms. 

Density is positively associated with entry, as expected. Adding it to the model reduces the source--GBB interaction from 0.0173 to 0.0110, but the interaction remains positive and statistically significant ($p=0.012$).\footnote{Note that this is not due to the somewhat different sample: the interaction term is 0.0173 on both the full sample from the main text and the reduced sample here.} The combined model also fits the data better than either model on its own (within $R^2$ of 0.031, against 0.019 and 0.021). We conclude that technological relatedness accounts for part of the original association, but the source--GBB interaction remains a significant predictor of green technology entry beyond relatedness.

\begin{table}[htbp]
\centering\small
\caption{Green entry with CPC relatedness density. All three models use the same observations and firm and target fixed effects. Brackets report 95\% confidence intervals from standard errors clustered by firm and target.}\label{tab:SI_entry_relatedness}
\begin{tabular}{lccc}
\toprule
Term & Density only & Source and GBB terms & Source and GBB terms + density \\
\midrule
GBB main effect & --- & 0.00791 & 0.01293 \\
 &  & $[-0.00600, 0.02182]$ & $[-0.00161, 0.02747]$ \\
Source main effect & --- & 0.01223 & 0.01004 \\
 &  & $[0.00366, 0.02080]$ & $[0.00212, 0.01797]$ \\
Source $\times$ GBB & --- & 0.01734 & 0.01099 \\
 &  & $[0.01110, 0.02357]$ & $[0.00257, 0.01941]$ \\
Relatedness density & 2.07181 & --- & 1.57595 \\
 & $[0.66538, 3.47824]$ &  & $[0.37301, 2.77889]$ \\
\midrule
Firm and target fixed effects & Yes & Yes & Yes \\
Observations & 578,417 & 578,417 & 578,417 \\
Within $R^2$ & 0.01890 & 0.02139 & 0.03139 \\
\bottomrule
\end{tabular}
\end{table}

\end{revision}

\section{Technological alliances}\label{sec:SI_alliances}

\subsection{GBB-based complementarity}\label{sec:SI_alliance_complementarity}
\begin{revision}
We focus on 5,947 focal firms and 7,190 candidate organizations with at least 500 patents. Organizations include not just firms but also universities, non-profit institutions and government organizations. Let $d_{fg}$ be the number of patents by firm $f$ in any source field connected to GBB $g$ in the source--GBB network and $a_{ig}$ the number of patents by organization $i$ in GBB $g$. An organization \emph{possesses} a GBB if it is specialized in the GBB (i.e., ${RCA}_{fg}^a\geq1$, see equation~\ref{eq:SI_entry_rca}) and has more than 10 patents in it: $A_{ig}=\mathbf{1}\{a_{ig}>10,\ {RCA}^a_{ig}\geq1\}$. Furthermore, we say that a firm \emph{needs} a GBB if it is specialized in source fields connected to that GBB and has more than 10 patents in these source fields: $R_{fg}=\mathbf{1}\{d_{fg}>10,\ RCA^d_{fg}\geq1\}$. For each pair of a firm $f$ and a candidate organization $o$, we now count the GBBs that $f$ needs but does not possess and that $o$ possesses:
\begin{equation}\label{eq:SI_complementarity}
K_{fo}=\sum_g R_{fg}(1-A_{fg})A_{og}.
\end{equation}
Note that $K_{fo}$ is directional: what $o$ can offer $f$ differs from what $f$ can offer $o$.

\end{revision}

\begin{figure}[htbp]
\centering
\includegraphics[width=0.9\textwidth]{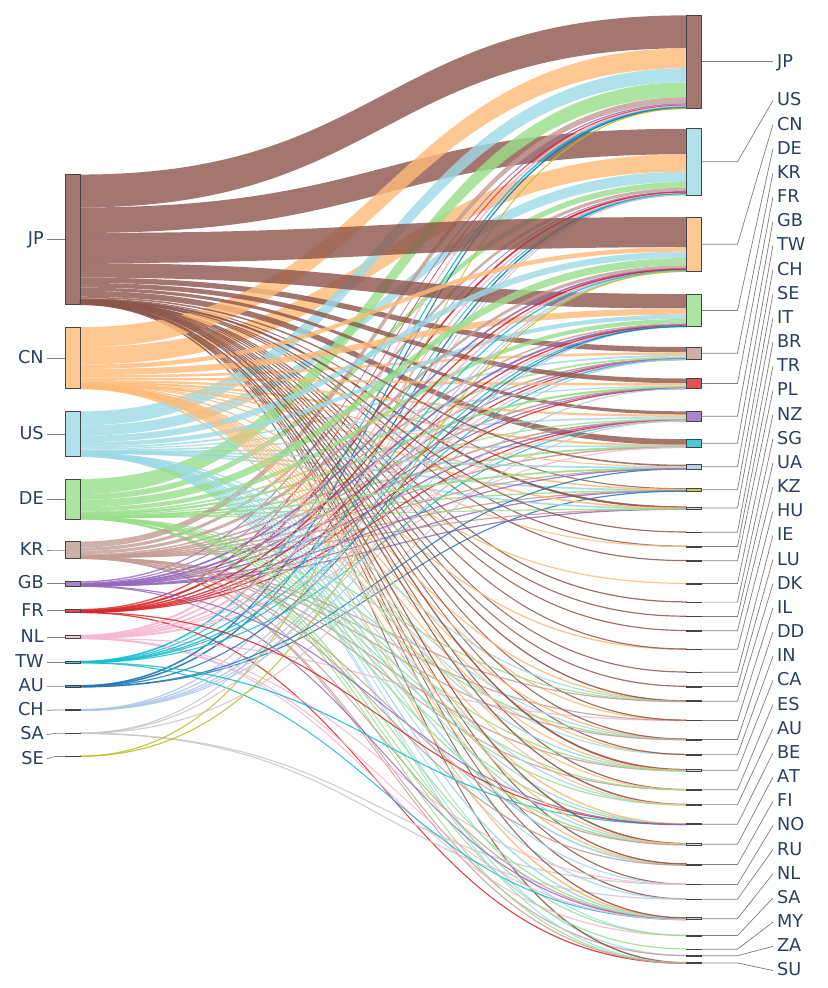}
\caption{Sankey diagram of most complementary candidates by country}
\label{fig:best_supplier_cntry}
\end{figure}

We can now identify the most complementary organization for each firm: the candidate with the highest $K_{fo}$. Ties are broken by the number of patents the candidate has in the GBBs it can supply to the firm. We next aggregate such most complementary pairs to the level of organization types and countries. The latter shows how often the most complementary candidate organization is located abroad. The resulting bipartite network is shown in Fig.~\ref{fig:best_supplier_cntry}. Table~\ref{tab:best_supplier_cntry} shows numerical values for firms in Germany, China and the US.

\begin{table}[htb]
\centering
\caption{Origins of most complementary candidate organizations for firms in the US, China and Germany}
\label{tab:best_supplier_cntry}
\small
\begin{tabular}{rrrr}
\toprule
\textbf{firm} & \textbf{organization} & \textbf{\# top candidates} & \textbf{ratio} \\
\midrule
DE                  & JP                  & 240              & 44.1\%         \\
DE                  & CN                  & 103              & 18.9\%         \\
DE                  & US                  & 72               & 13.2\%         \\
DE                  & DE                  & 70               & 12.9\%         \\
DE                  & KR                  & 27               & 5.0\%          \\
DE                  & GB                  & 15               & 2.8\%          \\
DE                  & NL                  & 5                & 0.9\%          \\
DE                  & FR                  & 4                & 0.7\%          \\
DE                  & CH                  & 4                & 0.7\%          \\
DE                  & TW                  & 3                & 0.6\%          \\
DE                  & AU                  & 1                & 0.2\%          \\
US                  & JP                  & 424              & 38.6\%         \\
US                  & CN                  & 290              & 26.4\%         \\
US                  & US                  & 171              & 15.6\%         \\
US                  & DE                  & 96               & 8.7\%          \\
US                  & KR                  & 54               & 4.9\%          \\
US                  & GB                  & 19               & 1.7\%          \\
US                  & FR                  & 17               & 1.5\%          \\
US                  & NL                  & 12               & 1.1\%          \\
US                  & AU                  & 10               & 0.9\%          \\
US                  & CH                  & 4                & 0.4\%          \\
US                  & TW                  & 1                & 0.1\%          \\
US                  & SA                  & 1                & 0.1\%          \\
CN                  & JP                  & 502              & 55.5\%         \\
CN                  & DE                  & 123              & 13.6\%         \\
CN                  & US                  & 106              & 11.7\%         \\
CN                  & CN                  & 74               & 8.2\%          \\
CN                  & KR                  & 53               & 5.9\%          \\
CN                  & NL                  & 14               & 1.5\%          \\
CN                  & GB                  & 12               & 1.3\%          \\
CN                  & FR                  & 7                & 0.8\%          \\
CN                  & CH                  & 5                & 0.6\%          \\
CN                  & TW                  & 5                & 0.6\%          \\
CN                  & SA                  & 2                & 0.2\%          \\
CN                  & SE                  & 1                & 0.1\%          \\
\bottomrule
\end{tabular}
\end{table}

\begin{revision}
\subsection{GBB complementarity and technological alliances}\label{sec:APP_allianceregression}

In the main text, we assessed the validity of our complementarity measure by exploring whether it can help predict technological alliances. We observe such alliances when patents are co-assigned to two or more organizations. Table~\ref{tab:pred_cogreenpat} reports the linear probability models used for these predictions in full. The dependent variable is one if a firm and a candidate organization are co-assigned at least one green patent and zero otherwise. The parameter estimates in this table thus describe the association between GBB complementarity and observed green patent co-assignment. The sample consists of 42,752,983 firm--candidate pairs, excluding pairs of a firm with itself.

Complementarity is positively associated with co-assignment across four different model specifications. Column (1) includes only $\log(1+K_{fo})$ as a regressor. Column (2) adds an indicator for whether a firm and candidate organization are located in the same country. Column (3) adds the logarithms of the number of patents of the firm and of the candidate organization. Column (4) replaces these by firm and candidate fixed effects, which absorb all time-invariant characteristics of either organization, yielding a complementarity coefficient of 0.00033 ($p<0.001$).

\begin{table}[htbp]
\centering
\caption{GBB complementarity and technological alliances: linear probability models.}
\label{tab:pred_cogreenpat}
\small
\setlength{\tabcolsep}{4pt}
\begin{tabular*}{\linewidth}{@{\extracolsep{\fill}}lcccc@{}}
\toprule
Dependent variable & \multicolumn{4}{c}{Green patent co-assignment (0/1)} \\
Model & (1) & (2) & (3) & (4) \\
\midrule
Constant & 0.000186$^{***}$ & -0.000251$^{***}$ & -0.011704$^{***}$ &  \\
 & (0.000021) & (0.000028) & (0.000891) &  \\
$\log(1+K_{fo})$ & 0.000674$^{***}$ & 0.000682$^{***}$ & 0.000301$^{***}$ & 0.000333$^{***}$ \\
 & (0.000050) & (0.000050) & (0.000029) & (0.000029) \\
Same country &  & 0.002898$^{***}$ & 0.002761$^{***}$ & 0.002806$^{***}$ \\
 &  & (0.000166) & (0.000153) & (0.000145) \\
Log candidate patents &  &  & 0.000753$^{***}$ &  \\
 &  &  & (0.000067) &  \\
Log firm patents &  &  & 0.000867$^{***}$ &  \\
 &  &  & (0.000072) &  \\
\midrule
Candidate fixed effects & No & No & No & Yes \\
Firm fixed effects & No & No & No & Yes \\
Observations & 42,752,983 & 42,752,983 & 42,752,983 & 42,752,983 \\
Positive observations & 25,525 & 25,525 & 25,525 & 25,525 \\
$R^2$ & 0.00035 & 0.00214 & 0.00418 & 0.01132 \\
Within $R^2$ & & & & 0.00157 \\
\bottomrule
\end{tabular*}
\par\smallskip
\begin{tabular}{p{.95\linewidth}}
\footnotesize Notes: Standard errors in parentheses are  clustered by firm and candidate organization. $^{***}p<0.001$. The sample contains all combinations of 5,947 firms and 7,190 candidate organizations, excluding pairs of a firm with itself.  \\
\end{tabular}
\end{table}

\subsection{Alliances versus co-assignments within corporate groups}\label{sec:APP_observed_samegroupaudit}

One concern is that our analysis conflates patent co-assignments to the different entities within the same corporation with technological alliances across different organizations. The fact that PATSTAT's \texttt{psn\_id}s already carefully disambiguate name-strings  mitigates this concern substantially. Nevertheless, the problem may persist in particular for multinational corporations, whose subsidiaries often have different names in different countries. Here we explore to what extent the consequent mistaking of within-corporation collaborations for technological alliances affects our results.

To do so, we take all co-assignments between two companies located in different countries. This yields 1,438 pairs of companies. We ask an LLM (DeepSeek-V4-Flash, with temperature set at 0) to assess, based on the two applicant names, organization types and countries, whether the two companies belong to the same corporate group. The LLM classifies each pair as \emph{same group} (one company is a subsidiary of the other, or both are subsidiaries of the same parent), \emph{different groups}, or \emph{uncertain}. Table~\ref{tab:observed_samegroupaudit} reports the results: 994 pairs (69\%) are classified as belonging to different groups, 262 (18\%) as belonging to the same group and 182 (13\%) as uncertain. 

\begin{table}[htbp]
\centering
\caption{LLM classification of co-assignments between companies in different countries.}
\label{tab:observed_samegroupaudit}
\small
\begin{tabular}{lrr}
\toprule
 & \# pairs & Share (\%) \\
\midrule
Different groups & 994 & 69.1 \\
Same group & 262 & 18.2 \\
Uncertain & 182 & 12.7 \\
\midrule
Total & 1,438 & 100.0 \\
\bottomrule
\end{tabular}
\end{table}

Next, we repeat the fixed-effects regression of column (4) in Table~\ref{tab:pred_cogreenpat}, first dropping the 262 pairs classified as same-corporate-group entities and then additionally dropping the 182 uncertain cases. Table~\ref{tab:alliance_ownership_robustness} reports the results. The association between complementarity and co-assignment remains positive and statistically significant. The coefficient is all but unaffected (0.00032 instead of 0.00033) when dropping both categories.

\begin{table}[htbp]
\centering
\caption{Alliance regressions after excluding potential within-corporate-group co-assignments. All models are based on the fixed-effects model of Table~\ref{tab:pred_cogreenpat}, column (4). 95\% confidence intervals in brackets.}
\label{tab:alliance_ownership_robustness}
\small
\begin{tabular}{lccc}
\toprule
 & Baseline & Dropping same group & Dropping same group and uncertain \\
\midrule
$\log(1+K_{fo})$ & 0.000333 & 0.000326 & 0.000318 \\
 & [0.000275, 0.000390] & [0.000269, 0.000383] & [0.000262, 0.000375] \\
\# observations & 42,752,983 & 42,752,459 & 42,752,095 \\
\# co-assignments & 25,525 & 25,001 & 24,637 \\
\bottomrule
\end{tabular}
\end{table}

\end{revision}

\begin{revision}

Finally, we address the same concern for the data presented in the Sankey diagram of Fig.~\ref{fig:best_supplier_cntry}. To do so, we screen the 5,947 pairs of a firm and its most complementary candidate for potential corporate-group linkages. To do so, we use the same LLM classification as before, which now classifies 22 pairs (0.4\%, of which 5 cross-border) as belonging to the same corporate group and 88 (1.5\%, of which 32 cross-border) as uncertain. The most complementary candidates are thus almost never part of the firm's own corporate group.

\end{revision}

\end{document}

%% file: tab/tab1.tex
\begin{tabular}{lcccc}
   \toprule
   Dep Var: & \multicolumn{4}{c}{$1(N_{f,c,1}>0)$}\\
   Model:                           & (1)            & (2)            & (3)            & (4)\\  
   \midrule
   \emph{Variables}\\
   $log(N_{f,src(c),0}+1)$                       & 0.0396$^{***}$ &                & 0.0143$^{***}$ & 0.0115$^{***}$\\   
                                    & (0.0051)       &                & (0.0039)       & (0.0042)\\   
   $log(N_{f,GBB(c),0}+1)$                      &                & 0.0497$^{***}$ & 0.0448$^{***}$ & 0.0079\\   
                                    &                & (0.0042)       & (0.0037)       & (0.0069)\\   
   $log(N_{f,src(c),0}+1)$   &                &                &                & 0.0173$^{***}$\\   
   $\times log(N_{f,GBB(c),0}+1)$            &                &                &                & (0.0031)\\   
   \midrule
   \emph{Fixed-effects}\\
   firm                       & Yes            & Yes            & Yes            & Yes\\  
   CCMT                              & Yes            & Yes            & Yes            & Yes\\  
   \midrule
   \emph{Fit statistics}\\
   Observations                     & 600,571        & 600,571        & 600,571        & 600,571\\  
   R$^2$                            & 0.15638        & 0.16600        & 0.16663        & 0.16780\\  
   Within R$^2$                     & 0.00704        & 0.01837        & 0.01911        & 0.02049\\  
   \bottomrule
   \multicolumn{5}{l}{\emph{Clustered (company $f$ \& CCMT $c$) standard-errors in parentheses}}\\
   \multicolumn{5}{l}{\emph{p-value: ***: 0.01, **: 0.05, *: 0.1}}\\
\end{tabular}

%% file: tab/tab_firm.tex
\begin{tabular}{cccc}
   \toprule
   Rank & BMW            & Tesla            & BYD\\  
   \midrule
   1  & Gas Storage and Transfer  & Advanced Solar Heat Collection                   & Modern Dynamo-Electric Machines\\ 
   2  & Advanced Welding Technologies        & Modern Dynamo-Electric Machines     & Advanced Semiconductor Imaging Technologies\\ 
   3  & Automotive Mechanisms Cluster       & Electric Motor Control Technologies                   & Advanced Solar Heat Collection\\ 
   4  & Electrochemical Energy Storage       & Electrical Coupling Devices  & \textcolor{red}{Electrochemical Energy Storage}\\
   5  & Advanced Engine Systems      & Electrochemical Energy Storage              & Advanced Ceramic Technologies
\\
   \bottomrule
\end{tabular}

%% file: tab/tab_cntry.tex
\begin{tabular}{cccc}
   \toprule
   Rank & Germany            & United States            & China\\  
   \midrule
   1  & Advanced Gasification  & Advanced Hydrocarbon Processing                   & Advanced Concrete Materials\\ 
   2  & Automotive Mechanisms Cluster & Hydrocarbon Catalysis                 &  Network Management and Communications\\ 
   3  & Advanced Engine Systems      & Advanced Medical Devices                   & Advanced Solar Heat Collection\\ 
   4  & Fluid Control Systems       & Molecular Sieves                            & Computer Security Systems\\
   5  & Advanced Steam Systems      & Biotech Processes and Compounds              & Advanced Lighting Technologies
\\
   \bottomrule
\end{tabular}